\documentclass[iicol,sn-basic]{sn-jnl}
\usepackage{graphicx}%
\usepackage{multirow}%
\usepackage{bm}
\usepackage{array}
\usepackage{graphicx}
\usepackage{subcaption}
\usepackage{amssymb}
\usepackage{amsmath}
\usepackage{amssymb}
\usepackage{amsfonts}
\usepackage{amsthm}%
\usepackage[table]{xcolor}
\usepackage{mathrsfs}%
\usepackage{epsfig}
\usepackage[title]{appendix}%
\usepackage{xcolor}%
\usepackage{textcomp}%
\usepackage{manyfoot}%
\usepackage{booktabs}%
\usepackage{algorithm}%
\usepackage{algorithmicx}%
\usepackage{algpseudocode}%
\usepackage{listings}%
\theoremstyle{thmstyleone}%

\theoremstyle{thmstyletwo}%
\theoremstyle{thmstylethree}%

\usepackage{longtable,float}
\usepackage{pifont}
\usepackage{color}

\usepackage{pgfplots}
\pgfplotsset{compat=1.18}
\usepackage{hyperref}
\definecolor{Red}{RGB}{217,30,24}
\definecolor{Orange}{RGB}{230,126,34}
\definecolor{Yellow}{RGB}{249,191,59}
\definecolor{Green}{RGB}{46,204,113}
\definecolor{Cyan}{RGB}{25,181,254}
\definecolor{Blue}{RGB}{31,58,147}
\definecolor{Pink}{RGB}{246,36,140}
\definecolor{Purple}{RGB}{154,18,179}
\definecolor{Brown}{RGB}{150,65,27}
\definecolor{Gray}{RGB}{174,184,184}
\definecolor{White}{RGB}{255,255,255}
\definecolor{Black}{RGB}{51,51,51}
\definecolor{Red_visual1}{RGB}{255,98,98}
\definecolor{Green_visual1}{RGB}{168,208,141}
\definecolor{Red1}{RGB}{192,0,0}
\definecolor{Red2}{RGB}{255,69,0}
\definecolor{Blue1}{RGB}{99,149,236}
\definecolor{Gray1}{RGB}{230,230,230}
\begin{document}

\title[Article Title]{Unsupervised Hyperspectral Image Super-Resolution via Self-Supervised Modality Decoupling}

\author[1]{Songcheng Du}\email{dusongcheng@mail.nwpu.edu.cn}
\equalcont{These authors contributed equally to this work.}
\author[1]{\fnm{Yang} \sur{Zou}}\email{archerv2@mail.nwpu.edu.cn}
\equalcont{These authors contributed equally to this work.}
\author[1]{Zixu~Wang}\email{wangzixu0827@mail.nwpu.edu.cn}
\author[2]{Xingyuan~Li}\email{xingyuan\_lxy@163.com}
\author*[1]{Ying~Li}\email{lybyp@nwpu.edu.cn}
\author[3]{Changjing~Shang}\email{cns@aber.ac.uk}
\author[3]{Qiang~Shen}\email{qqs@aber.ac.uk}

\affil[1]{\orgdiv{School of Computer Science}, \orgname{Northwestern Polytechnical University}, \orgaddress{\city{Xi'an}, \postcode{100190}, \country{China}}}

\affil[2]{\orgdiv{School of Software Technology}, \orgname{Dalian University of Technology}, \orgaddress{\city{Dalian}, \postcode{116086}, \country{China}}}

\affil[3]{Institute of Mathematics, Physics and Computer Science, Aberystwyth University,  Aberystwyth SY23 3DB, U.K}

\abstract{
Fusion-based hyperspectral image super-resolution aims to fuse low-resolution hyperspectral images (LR-HSIs) and high-resolution multispectral images (HR-MSIs) to reconstruct high spatial and high spectral resolution images. Current methods typically apply direct fusion from the two modalities without effective supervision, leading to an incomplete perception of deep modality-complementary information and a limited understanding of inter-modality correlations. To address these issues, we propose a simple yet effective solution for unsupervised HMIF, revealing that modality decoupling is key to improving fusion performance. Specifically, we propose an end-to-end self-supervised Modality-Decoupled Spatial-Spectral Fusion (MossFuse) framework that decouples shared and complementary information across modalities and aggregates a concise representation of both LR-HSIs and HR-MSIs to reduce modality redundancy. Also, we introduce the subspace clustering loss as a clear guide to decouple modality-shared features from modality-complementary ones. Systematic experiments over multiple datasets demonstrate that our simple and effective approach consistently outperforms the existing HMIF methods while requiring considerably fewer parameters with reduced inference time. 
The source code is in \href{https://github.com/dusongcheng/MossFuse}{MossFuse}.
}


\maketitle

\begin{figure*}
	\centering
	\includegraphics[scale=0.49]{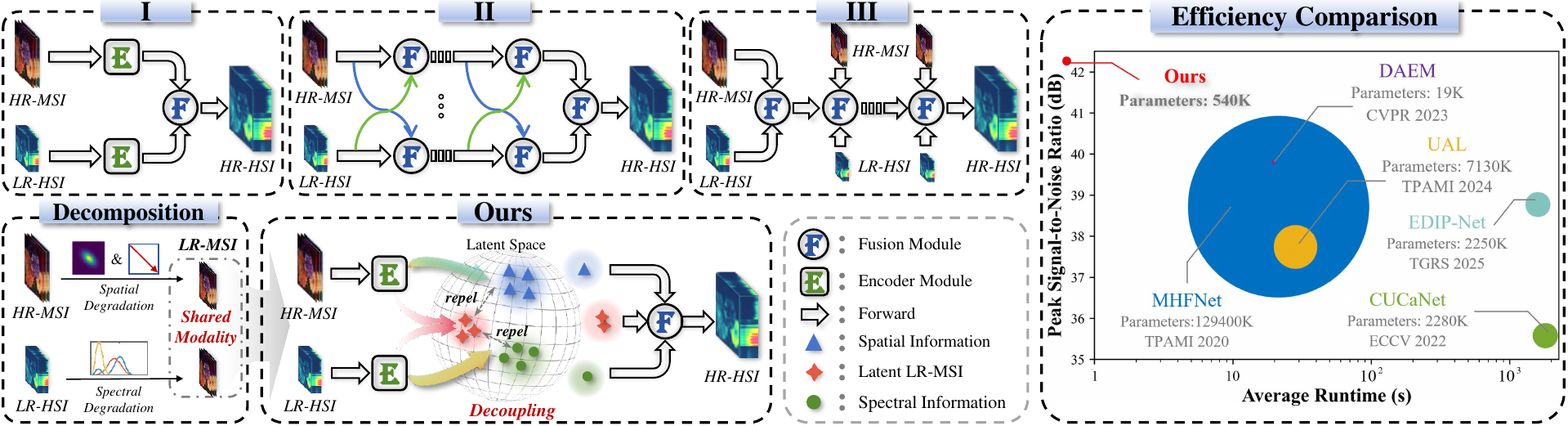}
	\caption{Algorithmic core difference with existing HMIF methods (\(\textbf{I}\), \(\textbf{II}\), and \(\textbf{III}\)) and efficiency comparison of our proposed MossFuse. We reveal that the modality decoupling is essential for unsupervised HMIF. Our simple solution significantly outperforms the existing HMIF methods while requiring considerably fewer parameters and reducing inference time.}
	\label{fig:teaser}
\end{figure*}

\section{Introduction}
\label{sec:intro}
Hyperspectral imaging achieves continuous spectral coverage with narrow sampling intervals (e.g., 10 nm), providing detailed reflectance signatures for precise material discrimination~\cite{yu2024unsupervised,liang2025dbmlla}. This unique capability makes it particularly valuable in domains such as environmental monitoring and earth observation~\cite{wang2023hyperspectral,lin2023metasurface,li2025frequency,zhang2023essaformer}. In the real world, however, optical systems typically offer either high spatial resolution with coarse spectral information (i.e., HR-MSI) or a greater number of spectral bands at lower spatial resolution (i.e., LR-HSI) due to the physical limitations of sensors~\cite{li2022drcr,li2022hasic,li2025enhanced}. To improve the usability of hyperspectral images (HSIs), the fusion of observed LR-HSI and HR-MSI has emerged as an efficient and promising method, drawing considerable attention~\cite{li2022deep1,li2022deep,wu2023hsr, yang2025deep}.

Without losing generality, HR-MSI and LR-HSI can be considered as the spatial and spectral degradation of a high-resolution hyperspectral image (HR-HSI), respectively~\cite{wang2020fusionnet,hu2024exploring}. The related observation models can be mathematically expressed as \( \text{Y} = \text{X}R \) and \( x = C\text{X} \), where $\text{X} \in \mathbb{R}^{H \times W \times B}$, $\text{Y} \in \mathbb{R}^{H \times W \times b}$ and $x \in \mathbb{R}^{h \times w \times B}$ represent the target HR-HSI, the HR-MSI and the LR-HSI, respectively.
Here, $H(h)$, $W(w)$, and $B(b)$ denote the height, width, and number of bands of the images, respectively, where $h\ll H$, $w\ll W$, $b\ll B$.
The matrix $R \in \mathbb{R}^{B\times b}$ represents the spectral response function (SRF), and $C \in \mathbb{R}^{hw\times HW}$ represents the point spread function (PSF), which is often modeled as a convolution with a blur kernel $K$ followed by a spatial downsampling operation.

Recently, numerous methods have been proposed to address the hyperspectral and multispectral image fusion (HMIF) problem~\cite{li2023x,gao2023enhanced}, which can be mainly categorized into three types, as illustrated in Fig.~\ref{fig:teaser}. In (I), the matrix factorization-based techniques facilitate information fusion through linear transformations~\cite{kawakami2011high,kanatsoulis2018hyperspectral}. In (II), the coupled branch architectures address the information deficiency in each modality by employing a multi-step cross-fusion process, which independently enhances both modalities~\cite{cucanet}. In (III), the repetitive integration approach combines observed LR-HSI and HR-MSI with a main branch to prevent information loss~\cite{dong2021model,mhfnet,li2023model,xu2025fusgat}. These architectures generally require the independent extraction of spatial and spectral feature representations from HR-MSI and LR-HSI prior to fusion. However, two major challenges persist: 1) Limited exploitation of the deep modality-complementary information, where the direct fusion of two modalities without effective supervision fails to adequately capture the deep modality-complementary information across modalities. This leads to redundant features and the deterioration of valuable modality-specific characteristics, ultimately yielding suboptimal fusion results. 2) Reduced efficiency, as the repeated extraction of redundant features increases computational costs. To overcome these inherent limitations, we develop an alternative approach that focuses on learning only the essential modality-complementary information through modality decoupling rather than relying on direct fusion. 

Specifically, as shown in Fig.~\ref{fig:teaser} (Decomposition), the LR-MSI can be obtained from both spatial degradation on the given HR-MSI \(\text{Y} \) and spectral degradation on the LR-HSI \(x\). Motivated by this degradation consistency observation, we note a key insight: HR-MSIs and LR-HSIs inherently contain both shared representations (LR-MSIs) and mutually exclusive complementary features in the latent space. In particular, we identify two distinct complementary components: 1) spatial high-frequency details lost during HR-MSI's spatial degradation process and 2) fine-grained spectral signatures discarded in LR-HSI's spectral dimensionality reduction. Building on this, we propose an end-to-end \textbf{Mo}dality-Decoupled \textbf{S}patial-\textbf{S}pectral Fusion (\textbf{MossFuse}) framework that reconstructs high-quality HR-HSIs with significantly fewer parameters and reduced inference time.

We first introduce a coarse-to-fine multi-modality decoupling process to decompose each given LR-HSI and HR-MSI into modality-complementary and modality-shared components within a latent space. To guide this decomposition, we devise a novel subspace clustering loss that effectively disentangles modality-shared and complementary information. Furthermore, we establish a self-supervised constraint mechanism that reinforces the decoupling process via a reversible transformation framework, enhancing the completeness and fidelity of the decoupled representations. Building on the inherent degradation properties of LR-HSIs and HR-MSIs, we introduce a dedicated degradation estimation module that accurately infers degradation parameters, significantly boosting the framework's robustness in unsupervised learning scenarios. Ultimately, we implement an adaptive modality aggregation strategy that effectively fuses the refined subspace representations, enabling accurate reconstruction of HR-HSIs.

In contrast to existing complex architectures, our work shows that effective fusion can be achieved through the integration of thoroughly decoupled modality-shared LR-MSIs and modality-complementary spatial and spectral information. This is achieved without relying on excessively deep networks or large parameter settings, yielding high-fidelity HR-HSIs with significantly improved efficiency in both computational complexity and inference speed. To this end, we propose MossFuse for unsupervised HMIF tasks, revealing the significance of modality decomposition. 

Our contributions are summarized as follows:

\begin{itemize}
\item We present a simple and effective solution for HMIF, demonstrating that effective fusion can be achieved through the integration of precisely decoupled modality components: the shared LR-MSI representation and complementary spatial and spectral information. To the best of our knowledge, this is the first work to address HMIF from a modality decomposition perspective.

% \item  We propose a simple and effective solution for HMIF, demonstrating that effective fusing the decoupled modality-shared LR-MSI and modality-complementary spatial and spectral information indeed enhances fusion performance. To the best of our knowledge, this is the first work to solve HMIF from a modality decomposition perspective.
% We propose a simple and effective solution for HMIF, demonstrating that modality decoupling is essential for hyperspectral and multispectral image fusion. To the best of our knowledge, this is the first work to solve HMIF from a modality decomposition perspective.

\item We introduce a specialized subspace clustering loss to enforce the decomposition process, effectively aggregating the shared LR-MSI components while discerning them from the modality-complementary spatial and spectral components.

\item We develop an end-to-end self-supervised modality-decoupled spatial-spectral fusion framework that decouples the shared and complementary information across modalities while aggregating the concise representation of the HSIs and MSIs to reduce the modality redundancy.

\item We present systematic experimental results validating our discovery and approach. Our simple and effective method outperforms the existing HMIF techniques across five popularly used datasets while requiring considerably fewer parameters with reduced inference time.

\end{itemize}

% \(x^{\da}\)

% \begin{figure*}
%   \centering
%   \begin{subfigure}{0.68\linewidth}
%     \fbox{\rule{0pt}{2in} \rule{.9\linewidth}{0pt}}
%     \caption{An example of a subfigure.}
%     \label{fig:short-a}
%   \end{subfigure}
%   \hfill
%   \begin{subfigure}{0.28\linewidth}
%     \fbox{\rule{0pt}{2in} \rule{.9\linewidth}{0pt}}
%     \caption{Another example of a subfigure.}
%     \label{fig:short-b}
%   \end{subfigure}
%   \caption{Example of a short caption, which should be centered.}
%   \label{fig:short}
% \end{figure*}

\section{Related Work}
\label{related works}

\subsection{Linear Decomposition based Fusion}
Fusion methods using linear decomposition typically decompose high-resolution multispectral images (HR-MSIs) and low-resolution hyperspectral images (LR-HSIs) to extract spatial and spectral features~\cite{han2020hyperspectral,fu2021fusion,li2023learning,ren2021locally,zou2024contourlet,li2025difiisr}. These methods can be broadly classified into matrix and tensor factorization paradigms~\cite{li2018fusing, xu2020hyperspectral, ye2023bayesian}. The matrix factorization method assumes that each pixel in an HSI is a linear combination of a finite set of reflectances. Kawakami et al.~\cite{kawakami2011high} initially applied an unmixing technique to estimate basis reflectance spectra from LR-HSI and subsequently used the coefficients obtained from HR-MSI to achieve the desired outcomes based on a sparse prior.
Yokoya et al.~\cite{yokoya2011coupled} introduced coupled non-negative matrix factorization (CNMF) to alternately estimate the spectral signatures and their corresponding coefficients. 
Unlike matrix factorization, tensor factorization directly decomposes an HR-HSI scene into a core tensor and dictionaries corresponding to the spatial and spectral dimensions. For instance, Dian et al.~\cite{dian2017hyperspectral} and Li et al.~\cite{li2018fusing} employed Tucker decomposition to capture non-local and coupled structural information, respectively. Kanatsoulis et al.~\cite{kanatsoulis2018hyperspectral} applied Canonical Polyadic (CP) decomposition to decompose the HR-HSI, estimating each factor matrix via least squares. 
Dian et al.~\cite{dian2023spectral} further integrate CP decomposition with deep neural networks by proposing an adaptive low-rank prior learning module, which leverages tensor factorization to enforce structured low-rank constraints while enabling efficient feature learning in a 1D spectral space.
Wu et al.~\cite{wu2024crodosr} proposed a cross-domain tensor singular value decomposition (t-SVD) framework to more effectively exploit the underlying low-rank structure of hyperspectral data.
Zhang et al.~\cite{zhang2018cluster} proposed a cluster-sparse field-based HSI reconstruction framework that explicitly models intrinsic priors in both spectral and spatial domains.
Fu et al.~\cite{fu2017adaptive} leveraged spectral correlations and spatial non-local self-similarity in degraded HSIs to learn an adaptive spatial-spectral dictionary.
Dian et al.~\cite{dian2024hyperspectral} introduce a generalized tensor nuclear norm (GTNN)-based fusion method that leverages clustering to enhance structural modeling of 3D HSI tensors.
Moreover, Akhtar et al.~\cite{akhtar2015bayesian} proposed a Bayesian sparse coding-based approach, where the HR-HSI is reconstructed by performing a linear combination of two key elements: a spectral dictionary, learned from the LR-HSI, and sparse coefficients, estimated from the HR-MSI.
In general, the effectiveness of such methods heavily relies on the accuracy of prior knowledge about the observed scene. However, in real-world scenarios, multiple light interactions with surfaces complicate linear representations, causing the solution to diverge from the true one.

\subsection{Deep Learning based Fusion}
Deep learning techniques, known for their ability to automatically capture the underlying structure of data~\cite{li2024contourlet,li2023progressive}, have inspired neural network-based fusion methods~\cite{liu2024promptfusion, li2025ustc,li2024object, wang2024general,zou2024enhancing}. Like linear decomposition approaches, many recent deep learning-based networks use a dual-branch autoencoder architecture to separately extract spatial and spectral information~\cite{cucanet}. 
Qu et al.~\cite{qu2018unsupervised} pioneered a deep learning-based unsupervised fusion network that incorporates a Dirichlet distribution-induced layer within a multi-stage alternating optimization process. 
Yao et al.~\cite{cucanet} presented a network with improved interpretability by incorporating spectral unmixing theory~\cite{keshava2002spectral} into a coupled-autoencoder framework, using cross-fusion layers to exchange spatial and spectral information. 
Liu et al.~\cite{liu2025low} proposed a low-rank Transformer network for spatial-spectral fusion, which effectively exploits both the spatial details and spectral characteristics embedded in MSI and LR-HSI. 
In addition, Dian et al.~\cite{dian2023spectral} propose an imaging model-guided optimization framework, which integrates physical acquisition priors with learned representations to regularize the reconstruction process by exploiting the intrinsic spatial and spectral priors of HSI.
Further, to address the zero-shot hyperspectral fusion problem, Dian et al.~\cite{dian2023zero} propose a novel and accurate method for quantitatively estimating the spectral and spatial response functions of imaging sensors degradation priors that enable faithful reconstruction without requiring ground-truth data.
To reinforce adaptability to different samples, more recent methods incorporate Test-Time Adaptation (TTA)~\cite{wang2020tent}, fine-tuning the network on test data. Zhang et al.~\cite{ual} proposed an unsupervised adaptation learning (UAL) framework that first learned a general image using deep networks and then adapted it to a specific HSI. Guo et al.~\cite{daem} proposed a coordination optimization framework to estimate the desired HSIs and their degradation~\cite{du2023degradation}, along with a partial fine-tuning strategy to cut computational costs, despite being time-consuming and having relatively poor generalization. To investigate the correlation between spatial and spectral information, Wu et al.~\cite{wu2024unsupervised} proposed a dual-branch super-resolution (SR) framework for HR-HSI reconstruction, where the interaction between the two branches serves as mutual constraints to enhance reconstruction quality. Additionally, Liu et al. proposed the SSMSFuse net~\cite{liu2025ssmsfuse} that coupled information at multiple scales with a two-branch network (Spa-Net and Spe-Net) for spectral and spatial coupling.
Moreover, with the popularity of the Mamba structure in the field of computer vision, Zhu et al.~\cite{zhu2025mamba} and Xiao et al.~\cite{sinet} propose MCIFNet and SINet, respectively, which both model HSI/MSI fusion using continuous implicit networks and show robust long-range modeling.
More recently, with a focus on practical deployment,  Li et al.~\cite{EDIP,li2024model} introduced a zero-shot fusion strategy, which optimizes the network solely based on the observed HR-MSI and LR-HSI pair, without relying on additional training datasets.

\section{The Proposed Method}
\definecolor{Red}{RGB}{217,30,24}
\begin{figure*}
	\centering
	\includegraphics[scale=0.185]{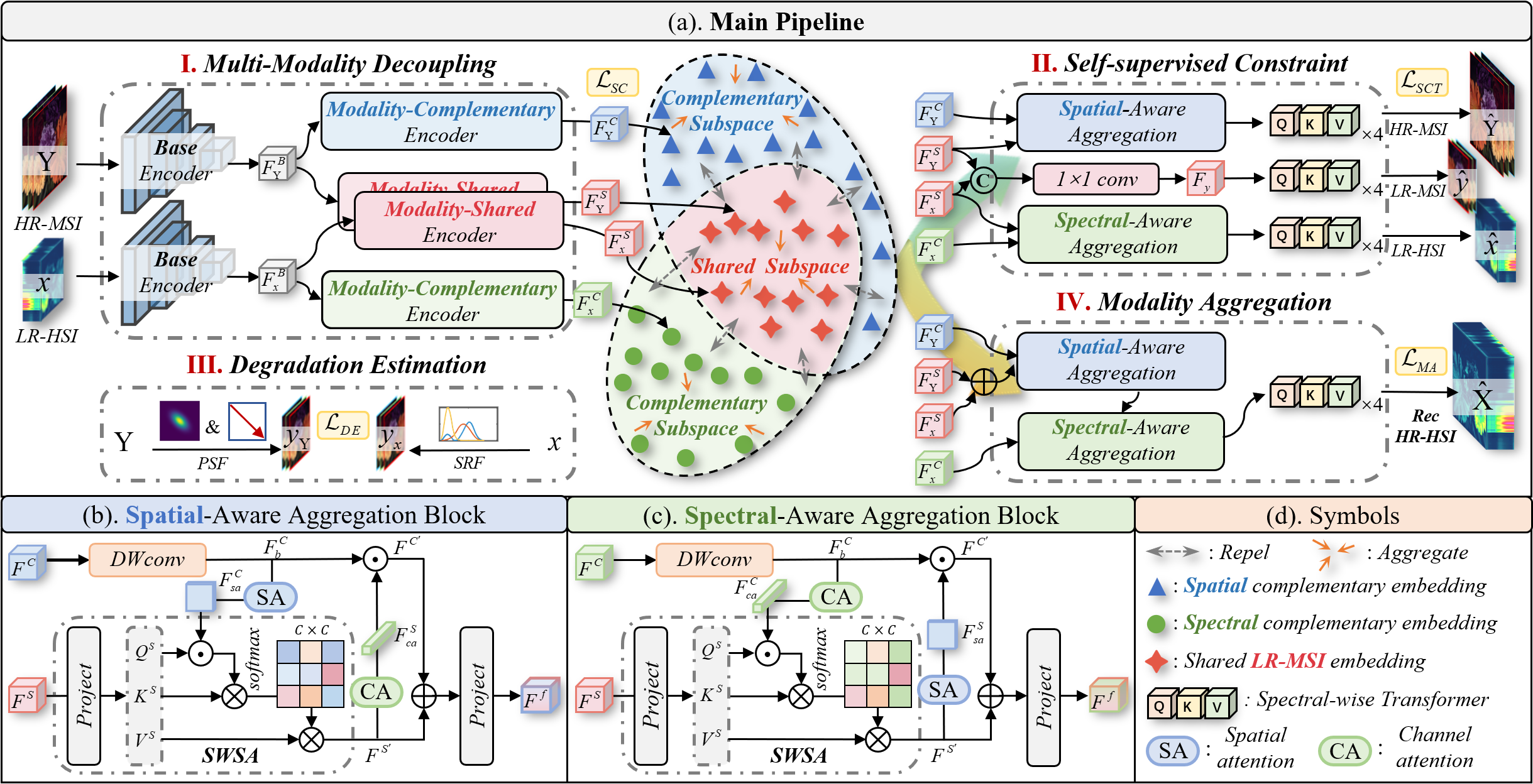}
	\caption{Architecture of MossFuse. \textbf{(a)} Main pipeline, consisting of four key components: \textcolor{Red1}{I.} Multi-Modality Decoupling, \textcolor{Red1}{II.} Self-supervised Constraint, \textcolor{Red1}{III.} Degradation Estimation, and \textcolor{Red1}{IV.} Modality Aggregation; with losses associated with each component being: $\mathcal{L}_{\text{SC}}$ (subspace clustering loss),  $\mathcal{L}_{\text{SCT}}$ (self-supervised constraint loss), $\mathcal{L}_{\text{DE}}$ (degradation estimation loss), and $\mathcal{L}_{\text{MA}}$ (modality aggregation loss). \textbf{(b)} and \textbf{(c)} Detailed architecture of Spatial- and Spectral-Aware Aggregation Blocks, respectively.}
	\label{fig_network}
\end{figure*}

\subsection{Motivation}
The fusion of a high-resolution multispectral image (HR-MSI) $\text{Y} \in \mathbb{R}^{H \times W \times b}$ and a low-resolution hyperspectral image (LR-HSI) $x \in \mathbb{R}^{h \times w \times B}$ aims to reconstruct the underlying high-resolution hyperspectral image (HR-HSI) $\text{X} \in \mathbb{R}^{H \times W \times B}$. The observed data are generated through a coupled distinct degradation processes:
\begin{equation}
    \text{Y} = D_\lambda(\text{X}), \quad x = D_s(\text{X}),
\end{equation}
where $D_\lambda: \mathbb{R}^B \to \mathbb{R}^b$ denotes spectral degradation (e.g., band integration via sensor spectral response functions), and $D_s: \mathbb{R}^{H \times W} \to \mathbb{R}^{h \times w}$ denotes spatial degradation (e.g., blur followed by downsampling).

A key observation arises from applying consistent degradations across modalities. Specifically, when we apply spatial degradation $D_s$ to $\text{Y}$ and spectral degradation $D_\lambda$ to $x$, both yield the same low-resolution multispectral image (LR-MSI):
\begin{equation}
\label{eq:degradation_consistency}
    y := D_s(\text{Y}) = D_s(D_\lambda(\text{X})) = D_\lambda(D_s(\text{X})) = D_\lambda(x),
\end{equation}
considering the commutativity of $D_s$ and $D_\lambda$~\cite{dian2023zero}. This consistency implies that $\text{Y}$ and $x$ share a common latent subspace corresponding precisely to $y$, which we define as the modality-shared representation, denoted $F^S$.

Beyond this shared component, each modality also contains unique, complementary information: The HR-MSI $\text{Y}$ provides high-frequency spatial details absent in $x$; The LR-HSI $x$ preserves fine-grained spectral signatures lost in $\text{Y}$.

We formalize these as modality-complementary representations, denoted $F_\text{Y}^C$ and $F_x^C$, respectively. Formally:
\begin{equation}
    \text{Y} \mapsto F^S \oplus F_\text{Y}^C, 
    x\mapsto F^S \oplus F_x^C,
\end{equation}
where the shared component $F^S$ acts as a structural bridge between modalities, while the complementary components encode discriminative attributes essential for faithful reconstruction.

% Recall that the degradation model of a hyperspectral image (HSI) can be formally formulated as:
% \begin{equation}
% \text{Y} = D_\lambda(\text{X}), x = D_s(\text{X}),
% \end{equation}
% where \(\text{X} \in \mathbb{R}^{H \times W \times B}\) represents the HR-HSI, $\text{Y} \in \mathbb{R}^{H \times W \times b}$ and $x \in \mathbb{R}^{h \times w \times B}$ represent the two inputs of the hyperspectral and multispectral image fusion (HMIF) task, HR-MSI and the LR-HSI, respectively. Here, \(D_\lambda:\mathbb{R}^B\rightarrow\mathbb{R}^b\) and \(D_s:\mathbb{R}^{H\times W}\rightarrow\mathbb{R}^{h\times w}\) represent the spectral- and spatial-wise degradation operators, implemented through spectral response functions and spatial blurring with downsampling operation.

% Further, we impose spatial degradation on \(\text{Y}\) (HR-MSI) and spectral degradation on \(x\) (LR-HSI), respectively, as follows:
% \begin{equation}
% y_{\text{Y}} = D_s(\text{Y}), y_x = D_\lambda(x),
% \end{equation}
% both \(\text{Y}\) and \(x\) converge into the identical low-resolution MSI (LR-MSI) domain (\(y_{\text{Y}}, y_x \in \mathbb{R}^{h \times w \times b}\)). 
% This degradation pathway reveals that the LR-MSI forms the shared latent subspace, bridging the HR-MSI and LR-HSI. Building on this degradation theoretical foundation, we establish a key observation: Both HR-MSI and LR-HSI can be decomposed into a modality-shared component (LR-MSI), along with modality-complementary components: spatial residual details from the HR-MSI and spectral residual information from the LR-HSI. 

However, current HMIF methods often rely on simplistic fusion strategies that directly merge different modalities, failing to capture deep cross-modality complementary relationships. This leads to two key issues: (1) feature redundancy from overlapping representations and (2) loss of modality-specific discriminative attributes, such as spectral fidelity or spatial sharpness. These shortcomings reduce fusion efficacy and output quality. To address this, we propose a modality-decoupling paradigm that selectively learns essential complementary features while eliminating redundant information.

In the following, we will elaborate on our methodology and implementation details.

\subsection{Method Overview}
As illustrated in Fig.~\ref{fig_network}(a), the main pipeline of our proposed modality-decoupled spatial-spectral fusion framework primarily integrates four key components: multi-modality decoupling, self-supervised constraint, degradation estimation, and modality aggregation, all of which work collaboratively to effectively generate high-quality spatial-spectral outputs.

Given paired inputs \(\text{Y}\) and \(x\), we first employ a multi-modality decoupling process to decouple them into modality-shared and modality-complementary components in the latent subspace, guided by the subspace clustering loss $\mathcal{L}_{\text{SC}}$. Then, to reinforce the completeness and fidelity of the decoupled representations, we implement a self-supervised learning strategy that incorporates spatial- and spectral-aware aggregation operations during the self-supervised constraint process. Furthermore, we estimate spatial and spectral degradation parameters based on the degradation models of HR-MSI and LR-HSI in the degradation estimation process. Finally, once the modality-shared and modality-complementary components are well separated and clustered, we reconstruct the desired HR-HSI \(\hat{\text{X}}\) by aggregating information from two input modalities through the modality aggregation process.

\subsection{Multi-Modality Decoupling}
We design a multi‑modality decoupling process to separate the LR‑HSI (\(x\)) and HR‑MSI (\(Y\)) into modality‑shared and modality‑complementary components within a latent space. First, both inputs are aligned spatially by upsampling \(x\) to match the size of $\text{Y}$. They are then passed through base encoders (comprising four spectral‑wise Transformer (SWT) blocks~\cite{zamir2022restormer}) to obtain shallow features $F^{B}_{\text{Y}}$ and $F^{B}_{x}$. 

Further, we design two couples of modality‑shared and modality‑complementary encoders to decouple the base features $F^{B}_{\text{Y}}$ and $F^{B}_{x}$ into shared and complementary components. The shared encoders produce \( F^{S}_{\text{Y}} \) and \( F^{S}_{x} \), capturing information shared to both modalities, while the complementary encoders yield \( F^{C}_{\text{Y}} \) (spatially complementary details) and \( F^{C}_{x} \) (spectrally complementary signatures). Formally, this decomposition can be expressed as:
\begin{equation}
F^{B}_{\text{Y}} = F^{S}_{\text{Y}} \oplus F^{C}_{\text{Y}}, F^{B}_{x} = F^{S}_{x} \oplus F^{C}_{x}.
\end{equation}

To extract spatially modality-complementary information, which captures degraded spatial details with up to ×32 resolution differences between LR-MSI and HR-MSI, we design the spatial modality-complementary encoder with three Large Kernel CNN (LK-CNN) blocks~\cite{zhang2024scaling}. These blocks leverage innovative kernel configurations to balance feature extraction efficiency and extended receptive field capture. 
For spectrally modality-complementary information, which contains high-resolution spectral cues absent in HR-MSI, we use an SWT block to extract residual spectral details from LR-HSI ($x$). 
To capture the modality-shared information, which reflects the common latent structure across both inputs—spatially degraded HR-MSI and spectrally degraded LR-HSI, we adopt modality-shared encoders that integrate both LK-CNN and SWT blocks. This joint structure is designed to extract spatial and spectral attributes simultaneously, enabling accurate alignment in the shared latent LR-MSI space.

% Together, these three components (shared, spatial-complementary, and spectral-complementary) form a complete and disentangled representation of the multi-modality inputs.

\subsection{Self-supervised Constraint}
Considering simply decoupling and clustering of the subspace features decoupled from different modalities may compromise the completeness and fidelity of the resulting representations. To address this, we introduce a self-supervised learning strategy that leverages feature aggregation operators to preserve both the integrity and discriminative quality of the decoupled subspace features.

Specifically, to constrain the two modality-shared representations \( F^{S}_{\text{Y}} \) and \( F^{S}_{x} \), we concatenate them channel-wise and apply a \( 1 \times 1 \) convolution, yielding the representation \( F_{y} \) of the LR-MSI. 

Also, we employ the spatial-aware aggregation block $\Phi^{spa}_{\theta_{\text{Y}}}(\cdot, \cdot)$ and the spectral-aware aggregation block $\Phi^{spe}_{\theta_{x}}(\cdot, \cdot)$ to recover the original HR-MSI \(\text{Y}\) and LR-HSI \(x\) from the decoupled representations [\(F^{C}_{\text{Y}}\), $F^{S}_{\text{Y}}$], and [$F^{C}_{x}$, $F^{S}_{x}$], respectively. Formally:
\begin{equation}
    F_{\text{Y}} = \Phi^{spa}_{\theta_{\text{Y}}}(F^{C}_{\text{Y}}, F^{S}_{\text{Y}}), \quad F_{x} = \Phi^{spe}_{\theta_{x}}(F^{C}_{x}, F^{S}_{x}).
\end{equation}
Finally, the three aggregated representations are decoded by their respective decoders \( \mathcal{D}_{\theta_{i}}(\cdot) \), \(i \in\{\text{Y},x,y\}\), and the reconstructed HR-MSI(\(\hat{\text{Y}}\)), LR-HSI(\(\hat{x}\)), and LR-MSI(\(\hat{y}\)) can be formulated as follows:
\begin{equation}
    \hat{\text{Y}} = \mathcal{D}_{\theta_{\text{Y}}}(F_{\text{\text{Y}}}),\hat{x} = \mathcal{D}_{\theta_{x}}(F_{x}),\hat{y} = \mathcal{D}_{\theta_{y}}(F_{y}).
\end{equation}

Thus, the decoupled process can be well constrained with the two inputs as references. More details are provided in Section \ref{Optimization}.

Note that both the spatial- and spectral-aware aggregation blocks are designed with two parallel branches: a convolutional branch that injects the modality-complementary representations and a spectral-wise self-attention (SWSA) branch that injects the modality-shared representations. These branches are integrated via a cross-external gated attention mechanism, which adaptively modulates the fusion of two types of features by dynamically adjusting feature importance across spatial and channel dimensions, respectively, enabling efficient and comprehensive feature fusion. The details of these two blocks are as follows:

\noindent \textbf{Spatial-Aware Aggregation Block.}
As illustrated in Fig.~\ref{fig_network}(b), our dual-branch block operates on two inputs: the modality-shared feature \( F^{S} \) and the complementary spatial feature \( F^{C}_{\text{Y}} \). The SWSA branch generates query-key-value triplets (\( Q^{S} \),\( K^{S} \),\( V^{S} \)) through linear projections of \( F^{S} \), while the convolutional branch processes \( F^{C}_{\text{Y}} \) with depth-wise convolution (DW-Conv) to produce base feature \( F^{C}_{b} \). To enable feature interaction, we employ a spatial attention (SA) mechanism~\cite{hu2018squeeze, chen2023dual} to extract an attention map \( F^{C}_{sa} \in \mathbb{R}^{H \times W \times 1} \) from \( F^{C}_{b} \). This spatial guidance is then fused with the SWSA branch through element-wise multiplication: 
\begin{equation}
    Q^{SC} = SA(F_b^C) \odot Q^S,
\end{equation}
where \( Q^{SC} \) represents the spatially refined query item, \(SA(\cdot)\) denotes the spatial attention operation and \(\odot\) denotes the element-wise multiplication operation. The final output of the SWSA module is computed via scaled element-wise multiplication attention with learnable temperature \( d \). Formally:
\begin{equation}
F^{S'} =  \text{softmax} \left(\frac{Q^{SC} (K^S)^T}{d}\right) V^S.
\end{equation}

For comprehensive feature integration, the transformer branch generates channel guidance through a channel attention (CA) mechanism~\cite{hu2018squeeze}. Particularly, we derive a channel attention map \( F^{S}_{ca} \in \mathbb{R}^{1 \times 1 \times C} \) from the enhanced feature \( F^{S^{\prime}} \), where \(C\) denotes the channel dimension. This attention map is then leveraged to recalibrate the convolutional branch via channel-wise multiplication dynamically: \(F^{C^{\prime}} = CA(F^{S^{\prime}}) \odot F^{C}_{b}\), where \(CA(\cdot)\) denotes the channel attention operation. The final fused representation \( F^{f} \) integrates cross-dimension information through residual summation followed by linear projection: \(F^{f} = Linear(F^{S^{\prime}}+ F^{C^{\prime}})\).

\noindent \textbf{Spectral-Aware Aggregation Block.} As depicted in Fig.~\ref{fig_network}(c), our architecture maintains dual-branch design, consistent with the Spatial-Aware Aggregation Block: The transformer branch converts modality-shared feature \( F^{S} \) into \( Q^{S} \), \( K^{S} \), and \( V^{S} \) via linear projections, while the convolutional branch processes modality-complementary spectral feature \(F^{C}_{\text{Y}}\) with DW-Conv to obtain spectral base feature \( F^{C}_{b} \).

To establish bidirectional spectral-spatial modulation, we first leverage channel attention (CA)~\cite{hu2018squeeze} to extract spectral guidance from \( F^{C}_{b} \) and interact with the SWSA branch: 
\begin{equation}
Q^{SC} = CA(F_b^C) \odot Q^S, 
\end{equation}
where \( Q^{SC} \) denotes the spectrally refined query. To reinforce the convolutional branch, spatial attention (SA) is applied to the SWSA branch output \( F^{S^{\prime}}\), followed by interaction with the convolutional features \(F_b^C\). Formally: 
\begin{equation}
    F^{C'} = SA(F^{S'})\odot F_b^C,
\end{equation}
where \(F^{S'} =  \text{softmax} \left(\frac{Q^{SC} (K^S)^T}{d}\right)V^S.\)

The final fused feature combines both branches through residual summation and linear projection: 
\begin{equation}
    F^f = \text{Linear}(F^{S'} + F^{C'}).
\end{equation} 

% In addition, the detailed architecture of SWSA, \( SA(\cdot) \) and \( CA(\cdot) \) are given in the Supplementary Materials.

\noindent \subsection{Degradation Estimation}
To alleviate the ill-posed nature of unsupervised HMIF task, we establish physics-grounded degradation models based on the degradation representations of HR-MSIs and LR-HSIs.

\noindent \textbf{Spatial Degradation Estimation.}
% Numerous studies suggest that the spatial-wise degradation of real-world scenarios can be approximated with anisotropic Gaussian kernels (AGK)~\cite{wang2024generalized, gu2019blind, yue2022blind}.  Accordingly, we simulate the point spread function (PSF), denoted as \(C\), following this approach.
Consensus from recent blind super-resolution studies~\cite{wang2024generalized, gu2019blind} indicates that real-world spatial degradation can be effectively modeled by anisotropic Gaussian kernels (AGK). Based on this, we parameterize the point spread function (PSF) \(C\) through three parameters: Eigenvalues \(\lambda_1\) and \(\lambda_2\) controlling kernel elongation, and Rotation angle \(\theta_K\) determining orientation. The covariance matrix \(\Sigma \in \mathbb{R}^{2 \times 2}\) is constructed via eigenvalue decomposition:
\begin{equation}
    \Sigma = R(\theta_K) \cdot \text{diag}(\lambda_1, \lambda_2) \cdot R(\theta_K)^T,
\end{equation}
where \(R(\theta_K) = \begin{bmatrix} \cos\theta_K & -\sin\theta_K \\ \sin\theta_K & \cos\theta_K \end{bmatrix}\) is the rotation matrix. The final blur kernel \(K\) is then generated through the standard 2D Gaussian formulation \(K \propto \exp\left(-\frac{1}{2} \mathbf{x}^T \Sigma^{-1} \mathbf{x}\right)\).
% To be specific, an AGK is characterized by three parameters: $\lambda_1$ and $\lambda_2$, which represent the eigenvalues of the kernel, and $\theta_K$, the rotation angle. The covariance matrix of the kernel is calculated as follows:
% {\footnotesize
%     \[
%     \Sigma=\left[\begin{array}{@{}cc@{}}
%     \cos\theta_K & -\sin\theta_K \\
%     \sin\theta_K & \cos\theta_K
%     \end{array}\right]
%     \left[\begin{array}{@{}cc@{}}
%     \lambda_1 & 0 \\
%     0 & \lambda_2
%     \end{array}\right]
%     \left[\begin{array}{@{}cc@{}}
%     \cos\theta_K & \sin\theta_K \\
%     -\sin\theta_K & \cos\theta_K
%     \end{array}\right].
%     \]
% }
% \begin{equation}
%  \Sigma=\left[\begin{array}{@{}cc@{}}
%     \cos\theta_K & -\sin\theta_K \\
%     \sin\theta_K & \cos\theta_K
%     \end{array}\right]
%     \left[\begin{array}{@{}cc@{}}
%     \lambda_1 & 0 \\
%     0 & \lambda_2
%     \end{array}\right]
%     \left[\begin{array}{@{}cc@{}}
%     \cos\theta_K & \sin\theta_K \\
%     -\sin\theta_K & \cos\theta_K
%     \end{array}\right].
% \end{equation}

% Further, the AKG-based blur kernel can be expressed as \(K = F_K(\Sigma)\), where $F_K$ is a function that generates the values of a 2D kernel based on kernel parameters~\cite{yue2022blind}.
 
\noindent \textbf{Spectral Degradation Estimation.}
Considering spectral degradation, the spectral response function (SRF) characterizes the sensitivity of a sensor at specific wavelengths, effectively representing an integration operation across the spectral channel dimension. Due to the inherent complexity of material properties, accurately modeling SRFs using simple mathematical formulations remains a significant challenge, as highlighted in recent studies~\cite{liu2022model, lian2024physics, li2024model, EDIP}. Given that SRFs can be discretized and expressed as a weighted summation of spectral curves, we propose to model the SRF using a \( 1 \times 1 \) convolutional operation, which inherently incorporates various physical constraints.

Specifically, to ensure model robustness, we employ a clamped activation function during each iteration to maintain the non-negativity of parameters. Furthermore, a normalization operator is applied to enforce the sum-to-one constraint across the spectral dimension, ensuring the SRF adheres to the required physical properties.

\subsection{Modality Aggregation}
With the modality-shared and modality-complementary components effectively and accurately decoupled from the HR-MSI and LR-HSI, we proceed to the modality aggregation process to reconstruct the desired HR-HSI (\(\hat{\text{X}} \)) by integrating these components. Starting with the modality-shared information, we progressively incorporate complementary spatial and spectral information. This process refines cross-attention in the spatial- and spectral-aware aggregation blocks using gate-like mechanisms and residual connections.

Recall that $F^{S}_{\text{Y}}$ and $F^{S}_{x}$ represent the same modality information, we first combine these shared components and aggregate the spatial complementary information from $F^{C}_{\text{Y}}$ through the spatial-aware aggregation block, denoted as $\Phi^{spa}_{\theta_{\text{X}}}(\cdot, \cdot)$. Following this, the spectral complementary information $F^{C}_{x}$ is integrated using the spectral-modality aggregation block, denoted as $\Phi^{spe}_{\theta_{\text{X}}}(\cdot, \cdot)$. Thus, the latent feature of the HR-HSI can be formulated as:
\begin{equation}
    F_{\text{X}} = \Phi^{spe}_{\theta_{\text{X}}}(\Phi^{spa}_{\theta_{\text{X}}}(F^{S}_{\text{Y}}+F^{S}_{x}, F^{C}_{\text{Y}}), F^{C}_{x}).
\end{equation}

Ultimately, the desired HR-HSI (\( \hat{\text{X}} \)) is reconstructed via the decoder \( \mathcal{D}_{\theta_{\text{X}}}(\cdot) \), which incorporates four SWT blocks. Formally, this can be expressed as: \( \hat{\text{X}} = \mathcal{D}_{\theta_{\text{X}}}(F_{\text{X}}) \).

\begin{table*}
\centering
%\scriptsize 
% \vspace{-1cm}
% \renewcommand\arraystretch{1}
%\setlength{\tabcolsep}{0.5mm}
% \renewcommand{\arraystretch}{0.9}
\resizebox{\linewidth}{!}{\begin{tabular}{l|cccc|cccc|cccc} 
\toprule[1.2pt]
Datasets & \multicolumn{4}{c|}{CAVE~\cite{yasuma2010generalized}}     & \multicolumn{4}{c|}{Harvard~\cite{chakrabarti2011statistics}}  & \multicolumn{4}{c}{NTIRE2018~\cite{arad2018ntire}}  \\
\cmidrule{1-13}
Methods  & PSNR $\uparrow$  & SSIM $\uparrow$  & SAM $\downarrow$   & ERGAS $\downarrow$ & PSNR $\uparrow$  & SSIM $\uparrow$  & SAM $\downarrow$   & ERGAS $\downarrow$ & PSNR $\uparrow$  & SSIM $\uparrow$  & SAM $\downarrow$   & ERGAS $\downarrow$   \\ 
\cmidrule{1-13}
CSU~\cite{lanaras2015hyperspectral}& 29.33 & 0.852 & 16.23 & 1.37  & 36.72 & 0.946 & 11.82 & 0.84  & 27.75 & 0.864 & 13.20 & 2.26    \\
NSSR~\cite{dong2016hyperspectral}  & 28.36 & 0.910 & 11.88 & 1.82  & 31.23 & 0.890 & 12.52 & 1.54  & 33.03 & 0.843 & 11.43 & 2.76    \\
HySure~\cite{simoes2014convex}     & 27.25 & 0.858 & 16.72 & 4.21  & 35.36 & 0.876 & 8.23  & 0.86  & 33.75 & 0.857 & 14.84 & 1.86    \\
MHFNet~\cite{mhfnet}               & 38.45 & 0.970 & 6.88  & 1.87  & 38.86 & 0.933 & 6.72  & 0.74  & 43.44 & 0.989 & 1.78 & 0.23    \\
EDIP-Net\cite{EDIP}                & 38.77 & 0.977 & 6.81  & 1.01  & 40.12 & 0.965 & 6.79  & 0.79  & 40.89 & 0.979 & 2.07 & 0.46     \\
CUCaNet\cite{cucanet}              & 35.46 & 0.929 & 10.43 & 0.97  & 37.44 & 0.940 & 7.46  & 0.68  & 39.66 & 0.986 & 4.69 & 0.49    \\
UAL~\cite{ual}                     & 37.78 & 0.976 & 9.24  & 0.68  & 40.92 & 0.936 & 7.35  & 0.64  & 44.83 & 0.993 & 1.32 & 0.10    \\
DAEM~\cite{daem}                   & \underline{39.83} & \underline{0.978} & \underline{6.79}  & \underline{0.54}  & \underline{42.87} & \underline{0.975} & \underline{4.06}  & \underline{0.46}  & \underline{46.73} & \underline{0.996} & \underline{0.96} & \underline{0.05}    \\ 
\midrule
Ours           & \textbf{42.15} & \textbf{0.983} & \textbf{6.48}  & \textbf{0.42}  & \textbf{44.62} & \textbf{0.991} & \textbf{3.76}  & \textbf{0.39}  & \textbf{48.64} & \textbf{0.999} & \textbf{0.77}  &  \textbf{0.04}   \\
\bottomrule[1.2pt]
\end{tabular}}
\caption{Quantitative comparison of different methods on three synthetic datasets. The best results are in \textbf{bold}, while the second-best results are in \underline{underlined}.}
\vspace{-0.5cm}
\label{table_synthetic_compare}
\end{table*}

\subsection{Network Optimization}
\label{Optimization}
\noindent \textbf{Subspace Clustering Loss.} To guide the multi-modality decoupling process, we propose a subspace clustering loss $\mathcal{L}_\text{SC}$ that encourages the network to align modality‑shared components while simultaneously repelling them from the modality‑complementary spatial and spectral features. 

As defined earlier, the multi-modality decoupling process decomposes the HR-MSI (\(\text{Y}\)) and LR-HSI (\(x\)) into four components: a modality-complementary spatial component \( F^{C}_{\text{Y}} \), a modality-complementary spectral component \( F^{C}_{x} \) and two modality-shared components \( F^{S}_{\text{Y}} \) and \( F^{S}_{x} \). The contrastive subspace clustering loss $\mathcal{L}_\text{SC}$ is computed by: 
\begin{flalign}
\scalebox{0.98}{$
\mathcal{L}_{\text{SC}} = -\log \frac{f(F^{S}_{\text{Y}}, F^{S}_{x})}{\sum_{i \in \{\text{Y},x\}} f(F^{S}_{i}, F^{C}_{i}) + \sum_{m \in \{\text{S,C}\}} f(F^{m}_{\text{Y}}, F^{m}_{x})}
$},
\end{flalign}
% \begin{flalign}
% \scalebox{0.92}{$
% \mathcal{L}_{\text{SC}} = -\log \frac{f(F^{S}_{\text{Y}}, F^{S}_{x})}{\displaystyle\sum_{i \in \{\text{Y},x\}} f(F^{S}_{i}, F^{C}_{i}) + \displaystyle\sum_{m \in \{\text{S,C}\}} f(F^{m}_{\text{Y}}, F^{m}_{x})}
% $},
% \end{flalign}
where \(f(a, b)\) = \(\exp^{cos(a, b)}\) is the latent space similarity metric, measuring the cosine similarity loss between two features \(a\) and \(b\).

This design is motivated by subspace clustering theory~\cite{elhamifar2013sparse} and contrastive representation learning~\cite{wang2020understanding}, both of which effectively promote latent structure separation and alignment. By maximizing the affinity of \(F^{S}_{\text{Y}}\) and \(F^{S}_{x}\) and minimizing their affinity with modality‑complementary spaces, \( \mathcal{L}_{\text{SC}} \) encourages a clear decomposition of features into distinct latent subspaces. Through gradient descent optimization of \( \mathcal{L}_{\text{SC}} \), redundancy across modalities is significantly reduced, improving the fusion quality.

\noindent \textbf{Self-supervised Constraint Loss.}
We propose the self-supervised constraint loss \( \mathcal{L}_{\text{SCT}} \) to preserve the integrity and fidelity of the decoupled modality information. We first incorporate two inputs HR-MSI (\(\text{Y}\)) and LR-HSI (\(x\)) as supervisory items to form the \(\mathcal{L}_{\text{SCT1}}\), ensuring the reversibility of the separated modalities. Formally: 
\begin{equation}
    \mathcal{L}_{\text{SCT1}}=\|\text{Y} - \hat{\text{Y}}\|_1 + \|x - \hat{x} \|_1.
\end{equation}

Moreover, we apply \(\mathcal{L}_{\text{SCT2}}\) on the decoded LR-MSI $\hat{y}$ to guide the modality decoupling. Recall that the LR-MSI can be obtained from both spatial degradation on the HR-MSI \(\text{Y} \) and spectral degradation on the LR-HSI \(x\), denoted as \(y_{\text{Y}} \) and \(y_{x} \), respectively. Therefore, we can construct the \(\mathcal{L}_{\text{SCT2}}\), such that:
% toward LR-MSI degraded from the HR-MSI and the LR-HSI, formally:

% \begin{flalign}
% \scalebox{0.8}{$
% \mathcal{L}_{\text{SCT2}} 
%     =\|y_\text{Y} - \hat{y} \|_1 + \|y_x - \hat{y} \|_1
%     =\|C\text{Y} - \hat{y} \|_1 + \|xR - \hat{y} \|_1.
% $}
% \end{flalign}
\begin{equation}
\begin{split}
    \mathcal{L}_{\text{SCT2}} 
    &= \|y_\text{Y} - \hat{y} \|_1 + \|y_x - \hat{y} \|_1 \\
    &= \|C\text{Y} - \hat{y} \|_1 + \|xR - \hat{y} \|_1.
\end{split}
\end{equation}

% \begin{equation}
%     \mathcal{L}_{\text{SCT2}} 
%     =\|y_\text{Y} - \hat{y} \|_1 + \|y_x - \hat{y} \|_1
%     =\|C\text{Y} - \hat{y} \|_1 + \|xR - \hat{y} \|_1.
% \end{equation}
Thus, $\mathcal{L}_\text{SCT}$ can be obtained by the combination of the two terms: 
\begin{equation}
\mathcal{L}_\text{SCT}=\mathcal{L}_\text{SCT1}+\mathcal{L}_\text{SCT2}.
\end{equation}

% The modality constraint loss \( \mathcal{L}_{\text{MCT}} \) ensures the network preserves the modal fidelity and integrity of the decoupled spatial, spectral, and latent LR-MSI components, thereby validating the proposed hypothesis.

\noindent \textbf{Degradation Estimation Loss.} 
% Since the LR-MSI can be obtained either through spatial degradation of HR-MSI or through spectral degradation of LR-HSI. 
Based on the degradation consistency theory, the LR-MSI can be equivalently derived from either HR-MSI through spatial degradation or LR-HSI through spectral degradation, respectively: \(y_{\text{Y}} = C\text{Y}, y_x=xR\). 

To ensure reliable estimation of spatial and spectral degradation parameters, we introduce the degradation estimation loss \(\mathcal{L}_{\text{DE}}\):
\begin{equation}
    \mathcal{L}_{\text{DE}} = \|y_{\text{Y}}-y_x\|_1= \|C\text{Y} - xR\|_1.
\end{equation}
% The Degradation Estimation Loss \( \mathcal{L}_{DE} \) plays a crucial role in the implementation of our framework.

\noindent \textbf{Modality Aggregation Loss.} To aggregate the decoupled modality features and reconstruct the desired HR-HSI $\hat{\text{X}}$ in an unsupervised manner, we leverage the HSI degradation model along with the estimated spatial and spectral degradation parameters SRF (\(R\)) and PSF (\(C\)). The modality aggregation loss is defined as:
\begin{equation}
    \mathcal{L}_{\text{MA}}=\|\text{Y} - \hat{\text{X}}R \|_1 + \| x - C\hat{\text{X}} \|_1,
\end{equation}
where \( \|\cdot\| \) represents the L1 loss, which effectively preserves both spatial structure and spectral fidelity during reconstruction.
% The modality reconstruction loss promotes sufficient interaction between different modality information, enabling accurate HMIF tasks.

Overall, the total training loss of the MossFuse, comprising the aforementioned loss terms, can be formulated as:
\begin{equation}
    \mathcal{L}_{\text{total}} = \mathcal{L}_{\text{MA}}+\alpha_1\mathcal{L}_{\text{SC}}+\alpha_2\mathcal{L}_{\text{SCT}}+\alpha_3\mathcal{L}_{\text{DE}}.
    \label{equation_loss}
\end{equation}

% In addition, the full training pipeline of XXX framework is provided in Algorithm 1. It is worth noting that XXX retains only the Multi-Modality Decoupling process and the Modality Reconstruction process in the inference phase.

\section{Experiment}
\label{sec:experiment}

\begin{figure*}
	\centering
	\includegraphics[scale=0.186]{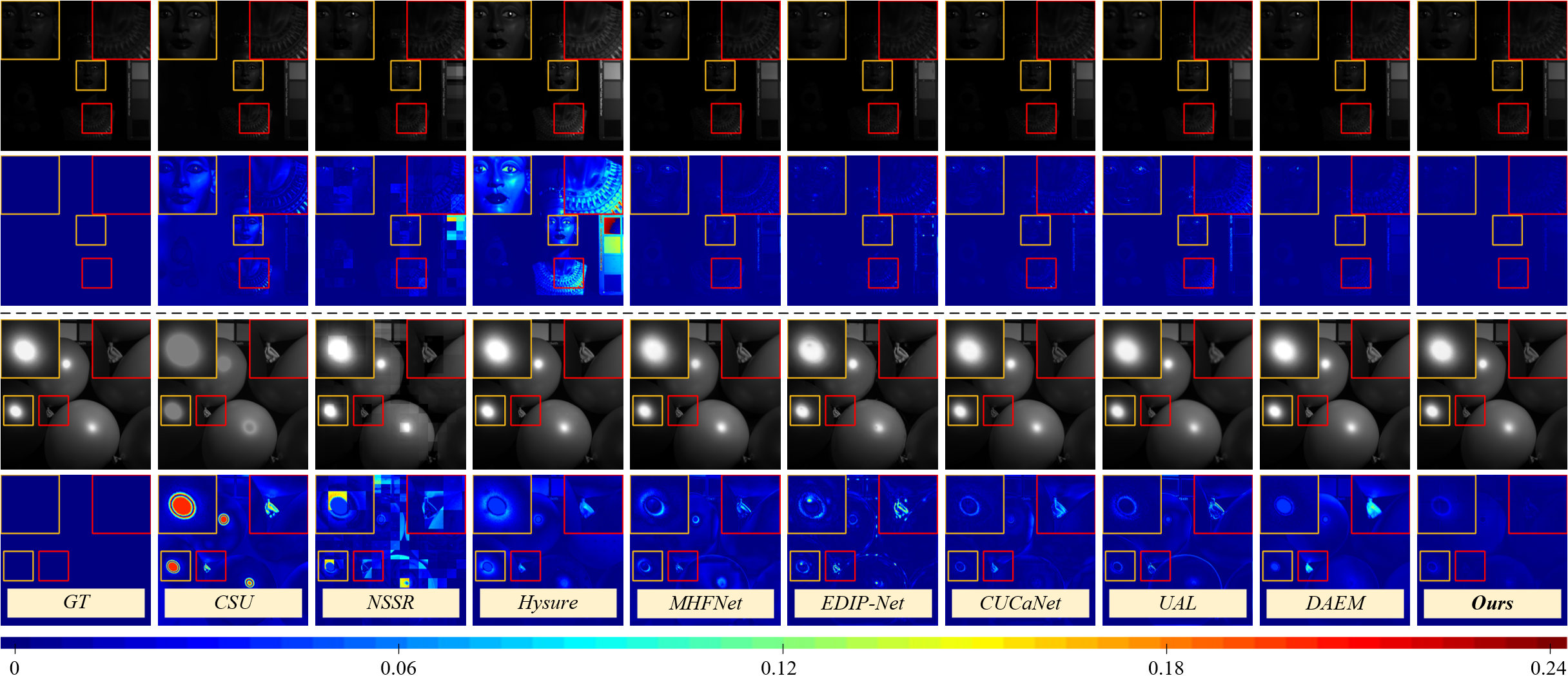}
	\caption{Visual reconstruction results and error images for 11th band of HR-HSI images from CAVE dataset. Zoomed-in regions are provided to highlight details.}
	\label{fig_cave}
\end{figure*}

\begin{figure*}
	\centering
	\includegraphics[scale=0.186]{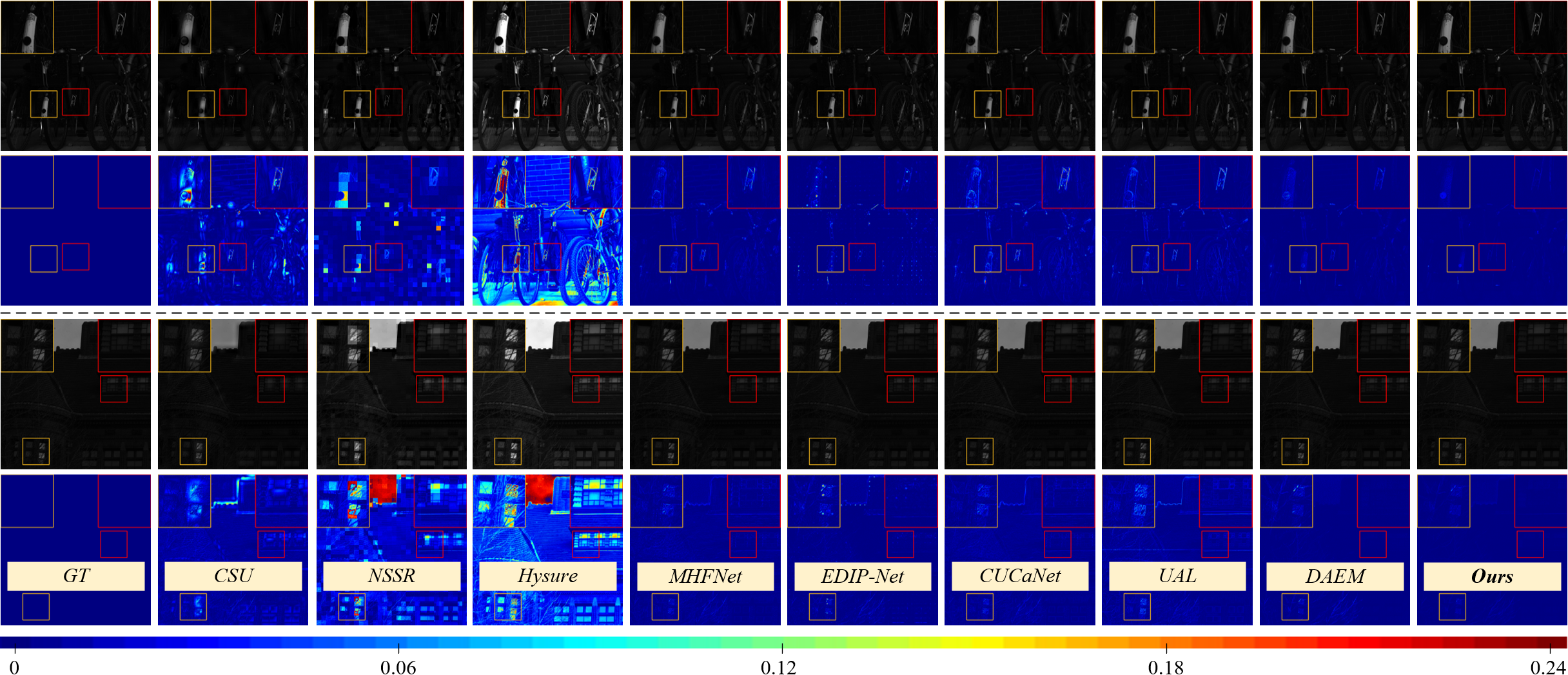}
	\caption{Visual reconstruction results and error images for the 11th band of the HR-HSI images from Harvard dataset. Zoomed-in regions are provided to highlight details.}
	\label{fig_harvard}
\end{figure*}

\begin{figure*}
	\centering
	\includegraphics[scale=0.186]{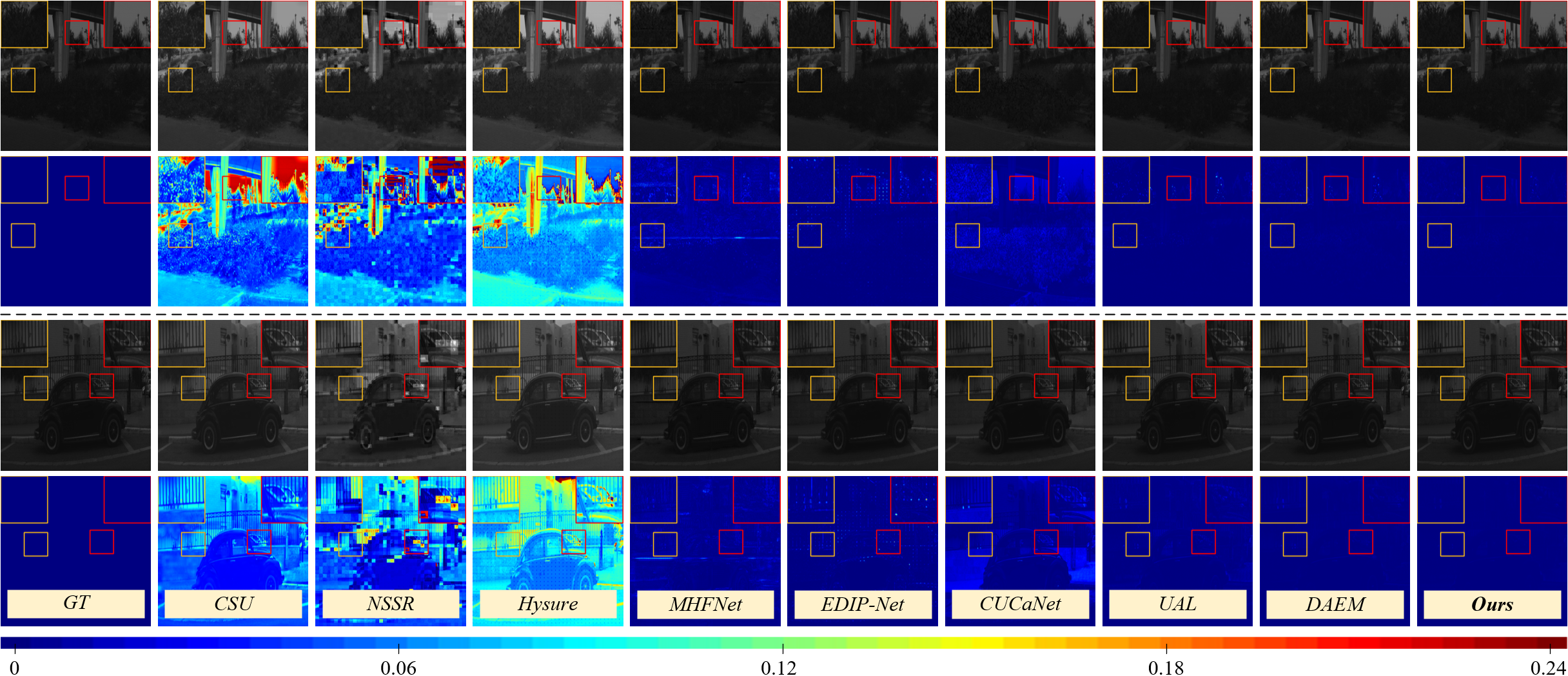}
	\caption{Visual reconstruction results and error images for 11th band of HR-HSI images from NTIRE2018 dataset. Please zoom in to see the details better.
	}
	\label{fig_ntire}
\end{figure*}

\begin{figure*}[ht]
    \centering
    \begin{subfigure}{0.33\textwidth}
        \includegraphics[width=\linewidth]{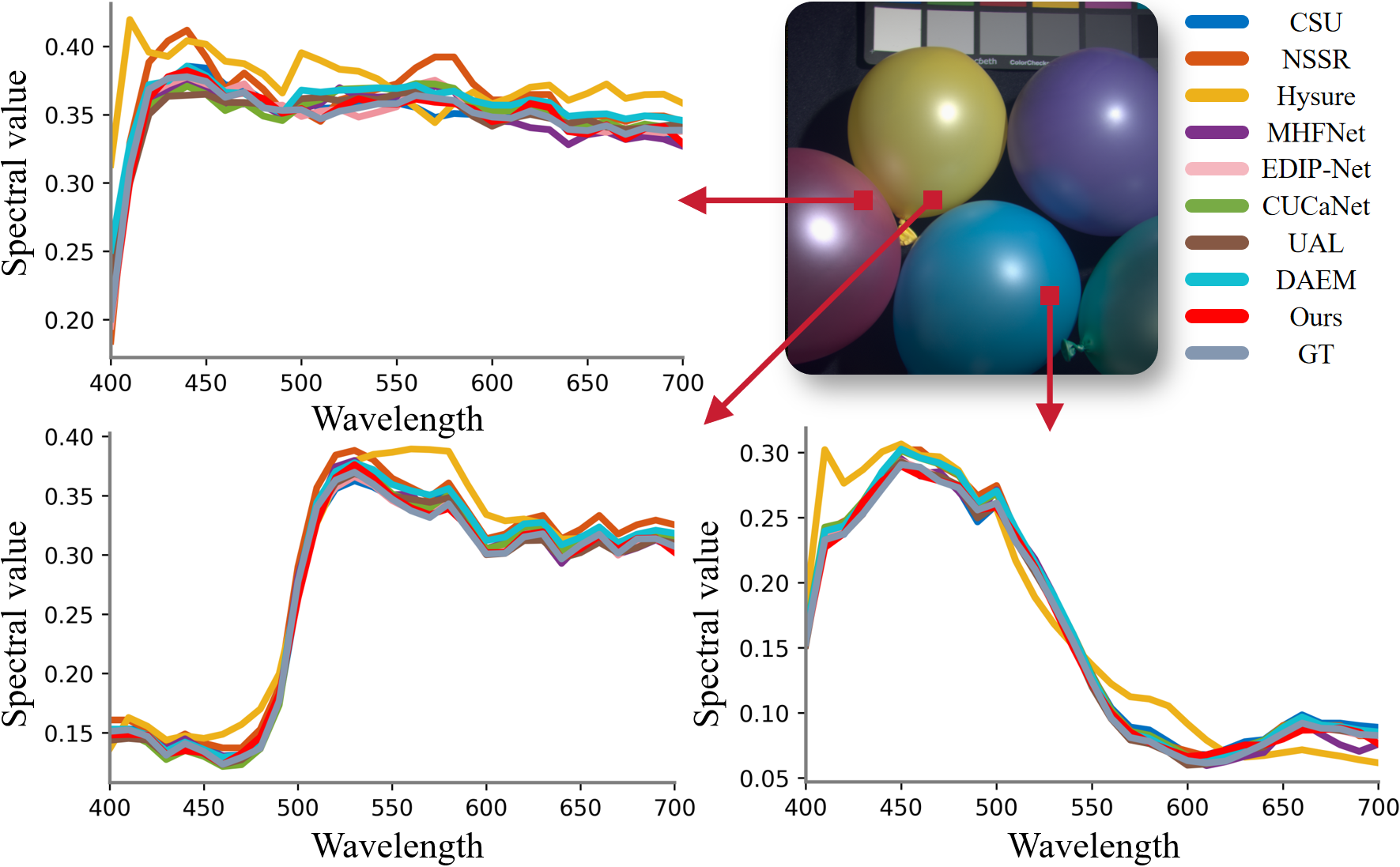}
        \caption{CAVE dataset}
        \label{fig:image1}
    \end{subfigure}%
    \hfill
    \begin{subfigure}{0.33\textwidth}
        \includegraphics[width=\linewidth]{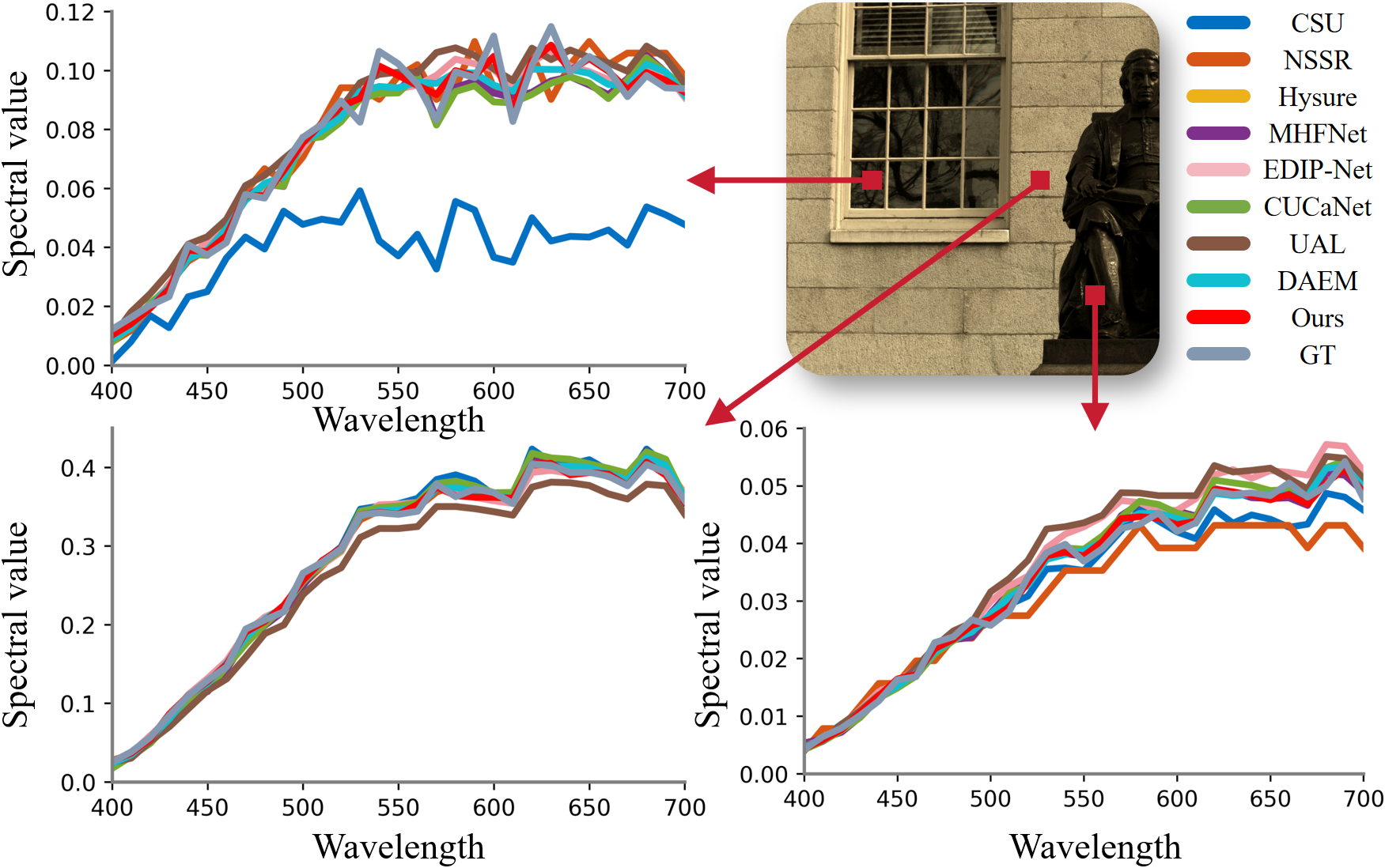}
        \caption{Harvard dataset}
        \label{fig:image2}
    \end{subfigure}%
    \hfill
    \begin{subfigure}{0.33\textwidth}
        \includegraphics[width=\linewidth]{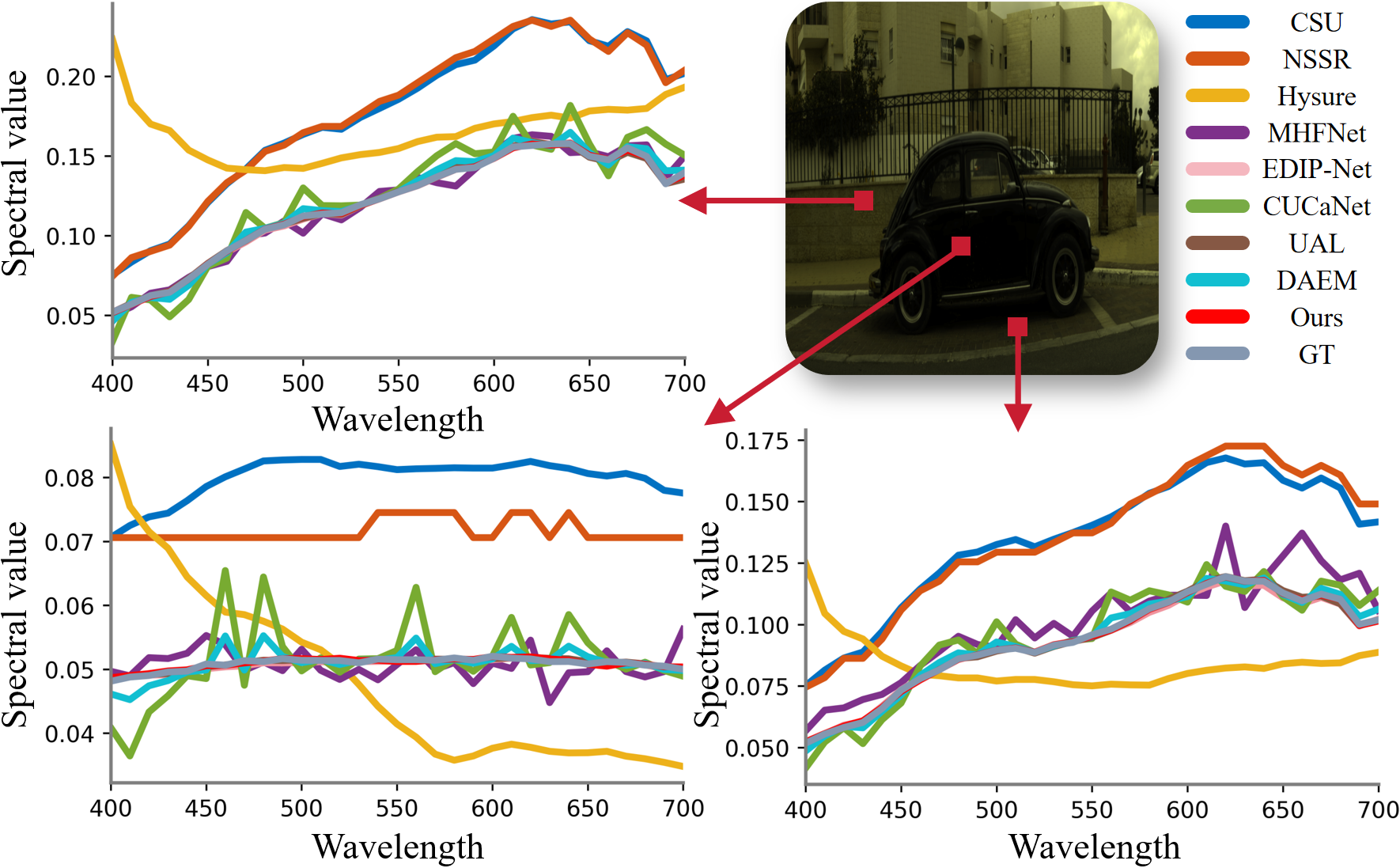}
        \caption{NTIRE2018 dataset}
        \label{fig:image3}
    \end{subfigure}
    \caption{Comparison of reconstructed spectra curves in (a) CAVE, (b) Harvard, and (c) NTIRE2018 datasets. Zoomed-in regions are provided to highlight details.}
    \label{fig_spectral}
\end{figure*}

\begin{figure*}
	\centering
	\includegraphics[scale=0.18]{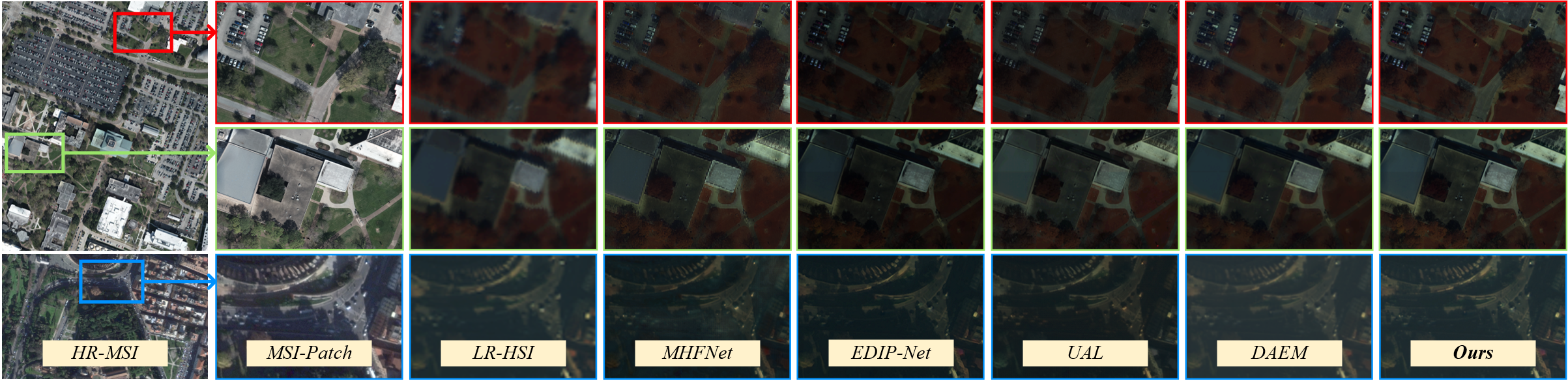}
	\caption{Visual reconstruction results from NCALM (upper part) and WV-2 (bottom part) datasets. We show images with 25-16-7 and 5-3-2 as R-G-B, respectively.}
	\label{fig_NCALM}
\end{figure*}

\subsection{Experiment Setup}
\noindent \textbf{Dataset.} 
In this study, we conducted comprehensive evaluations using a diverse collection of five commonly used benchmark datasets in this field, comprising three synthetic HSI datasets including CAVE~\cite{yasuma2010generalized}, Harvard~\cite{chakrabarti2011statistics}, and NTIRE2018~\cite{arad2018ntire}, as well as two real remote sensing HSI datasets, including National Center for Airborne Laser Mapping (NCALM)~\cite{xu2019advanced} and Worldview2 (WV-2)~\cite{signet}, for evaluation. 

Specifically, the CAVE dataset includes 32 indoor hyperspectral images, each containing 31 spectral bands covering the visible spectrum from 400 to 700 nm with a spatial resolution of 512 × 512 pixels. The Harvard dataset collects 50 outdoor HSI samples recorded under daylight illumination, each with a high spatial resolution of 1392 × 1040 pixels and comprising 31 spectral bands ranging from 420 nm to 720 nm at 10 nm intervals. The NTIRE 2018 dataset, from the first spectral image super-resolution challenge, includes 54 natural HSIs for training and 5 for testing, each with a spatial resolution of 1392 × 1300 pixels and 31 spectral bands from 400 nm to 700 nm at 10 nm intervals. The NCALM dataset, which was published by the 2018 IEEE GRSS Data Fusion Contest, features a 48-band HSI covering an extensive spectral range from 380 to 1050 nm with a spatial resolution of 1202 × 4172 pixels, accompanied by a 3-band MSI at 24040 × 83440 pixels. The WV-2 dataset, captured by a commercial satellite, includes an 8-band HSI with a spatial resolution of 418 × 685 pixels and a corresponding 3-band MSI at 1677 × 2633 pixels. 

For the synthetic datasets (CAVE, Harvard, and NTIRE 2018), we established the following training-testing splits: 20, 30, and 54 pairs of images for training, respectively, with the remainder used for testing. To generate corresponding HR-MSI and LR-HSI pairs, we employed a simulation process using the spectral response function (SRF) and point spread function (PSF) from~\cite{jiang2013space}. Specifically, the HR-MSI was simulated through spectral downsampling, while the LR-HSI was obtained by applying the selected PSF, followed by spatial downsampling with a scale factor of 32 to the HR-HSI.

For the real-world remote sensing datasets NCALM and WV-2, we adopted the standard evaluation protocol established in~\cite{signet}, utilizing the top half of both HSI and MSI images for training and the bottom half for testing. 
% Following the Wald protocol~\cite{wald2002data}, we generated the training pairs through systematic downsampling: the input images were downsampled with factors of 4 and 20 for NCALM and WV-2, respectively, to create corresponding LR-HSI and HR-MSI pairs.

\noindent \textbf{Compared Methods.} 
To demonstrate the superiority and generality of the proposed approach, we perform extensive comparisons with both traditional optimization-based and recent deep learning-based unsupervised fusion methods. 
Specifically, three representative tradition-based algorithms are included: CSU~\cite{lanaras2015hyperspectral}, NSSR~\cite{dong2016hyperspectral} and HySure~\cite{simoes2014convex}. In addition, five state-of-the-art (SOTA) unsupervised deep fusion networks are employed for comprehensive comparison, including: MHF~\cite{mhfnet}, CUCaNet~\cite{cucanet}, DEIP-Net~\cite{EDIP}, UAL~\cite{ual}, and DAEM~\cite{daem}.

% For comparison, we conducted comparative experiments against three tradition-based algorithms (including CSU~\cite{lanaras2015hyperspectral}, NSSR~\cite{dong2016hyperspectral}, HySure~\cite{simoes2014convex}) and five state-of-the-art (SOTA) unsupervised fusion algorithms (including MHF~\cite{mhfnet}, CUCaNet~\cite{cucanet}, DEIP-Net~\cite{EDIP}, UAL~\cite{ual}, and DAEM~\cite{daem}).

\noindent \textbf{Evaluation Measurements.}
To comprehensively assess the effectiveness and reliability of the proposed method, for the three simulated dataset, four quantitative metrics are adopted~\cite{signet}: the peak signal-to-noise ratio (PSNR), the structural similarity index (SSIM)~\cite{wang2004image}, the spectral angle mapper (SAM)~\cite{yuhas1992discrimination}, and the relative dimensionless global error in synthesis (ERGAS)~\cite{wald2000quality}. For the two real datasets, three no-reference metrics $D_\lambda$, $D_s$, and HQNR, are employed. 

\subsection{Implementation Details}
Our algorithm was implemented using the PyTorch framework, with all experiments conducted on an NVIDIA 4090 GPU. For optimizing the parameters, we utilized the Adam optimizer \cite{kingma2014adam} with settings $\beta_1=0.9$, $\beta_2=0.999$ and $\epsilon=10^{-8}$. To adjust the learning rate, we implemented a cosine annealing decay strategy, reducing it from $10^{-3}$ to $10^{-6}$ with the number of epochs set to 300. Moreover, to accommodate hardware constraints, we preprocessed the training samples by cropping them into 128 \(\times\) 128. Given the limited scale of available datasets, we applied comprehensive data augmentation techniques, including random rotation and flipping operations on all input pairs. The training process utilized a batch size of 2 across all three datasets to maintain consistency. As for the multi-term loss functions defined in Eq.\ref{equation_loss}, we carefully balanced the contribution of each component by setting the weighting coefficients \(\alpha_1\) to \(\alpha_3\) to 0.1,1, and 1, respectively. This configuration ensures that all loss terms maintain comparable orders of magnitude during optimization, thereby facilitating stable training convergence. 
Regarding feature dimensions throughout the network: the input HR-MSI \(\text{Y} \in \mathbb{R}^{H \times W \times b}\) and upsampled LR-HSI \(x \in \mathbb{R}^{h \times w \times B}\) are processed by two separate base encoders to generate initial features \(F^{B}_{\text{Y}},F^{B}_{x}\in \mathbb{R}^{H \times W \times C}\), where \(C=48\) is the hidden channel dimension. Moreover, all subsequent feature maps, including the decoupled modality-shared (\(F^{S}_{\text{Y}}\) and \(F^{S}_{x}\)) modality-complementary (\(F^{C}_{\text{Y}}\) and \(F^{C}_{x}\)) representations preserve the spatial resolution and dimensionality \(H \times W \times C\). Finally, the aggregated latent feature \(F^{C}_{\text{X}}\in \mathbb{R}^{H \times W \times C}\) is mapped to the target HR-HSI output \(\hat{\text{X}}\in \mathbb{R}^{H \times W \times B}\) in the decoder via channel reshaping.

\subsection{Performance Comparison}
\noindent \textbf{Synthetic Dataset.} 
Our comprehensive evaluation demonstrates the superior performance of MossFuse through both quantitative metrics and qualitative visual analysis across the three synthetic datasets.

The quantitative comparison, presented in Table~\ref{table_synthetic_compare}, reveals that MossFuse consistently outperforms all competing methods across three benchmark datasets, CAVE, Harvard, and NTIRE2018. 
On the CAVE dataset, our method establishes new performance records with a 2.32 dB PSNR improvement and a 0.31 SAM reduction compared to the strongest baseline, reflecting both enhanced reconstruction fidelity and exceptional spectral preservation. For the Harvard dataset, MossFuse achieves a PSNR of 44.62 (+1.75 dB over the second-best method) and an ERGAS score of 0.39, indicating statistically significant improvements in both spectral consistency and spatial accuracy. Notably, the NTIRE 2018 dataset evaluation reveals MossFuse's performance with a PSNR of 48.64 and near-perfect SSIM of 0.999—metrics that surpass competing approaches by substantial margins. These cross-dataset results systematically validate our method's robustness in maintaining critical hyperspectral features while minimizing reconstruction errors across diverse imaging scenarios.
Traditional methods, including CSU~\cite{lanaras2015hyperspectral} and HySure~\cite{simoes2014convex}, exhibit inherent limitations in modeling intricate spectral-spatial interdependencies, primarily stemming from their restricted representational capacity for high-dimensional feature learning.

For qualitative assessment, we provide detailed visual comparisons across all three synthetic datasets in Fig.~\ref{fig_cave}, Fig.~\ref{fig_harvard}, and Fig.~\ref{fig_ntire}. Each figure presents a dual-view analysis: the top row displays the reconstructed spectral band images, while the bottom row shows the corresponding reconstruction error heatmaps. Two strategically selected regions of interest are magnified for detailed comparison. The visual results clearly demonstrate that MossFuse generates reconstructions that are visually indistinguishable from the ground truth, consistently achieving the lowest reconstruction error across all test cases. 

To further assess the spectral fidelity of the reconstructed results, we conducted a detailed spectral signature analysis, as illustrated in Fig.~\ref{fig_spectral}. Specifically, we extract normalized spectral intensity curves from representative sample points and compare them across different methods. The results show that MossFuse achieves the closest alignment with the ground truth spectral signatures, exhibiting minimal distortion across wavelengths. This quantitative spectral analysis, combined with the visual results, provides compelling evidence of our method's effectiveness in preserving both spatial details and spectral accuracy, setting a new benchmark in fusion-based hyperspectral image super-resolution performance.

\begin{table}
\centering
%\scriptsize 
% \vspace{-1cm}
% \renewcommand\arraystretch{1}
%\setlength{\tabcolsep}{0.5mm}
% \renewcommand{\arraystretch}{0.9}
\resizebox{\linewidth}{!}{\begin{tabular}{l|ccc|ccc} 
\toprule[1.2pt]
Datasets & \multicolumn{3}{c|}{NCALM}     & \multicolumn{3}{c}{WV-2}  \\
\cmidrule{1-7}
Methods  & $D_\lambda$ $\downarrow$  & $D_s$ $\downarrow$  & HQRN $\uparrow$   & $D_\lambda$ $\downarrow$  & $D_s$ $\downarrow$  & HQRN $\uparrow$  \\
\cmidrule{1-7}
MHFNet~\cite{mhfnet}     & 0.056 & 0.042 & 0.904 & 0.036 & 0.048 & 0.918    \\
EDIP-Net\cite{EDIP}      & 0.047 & 0.028 & 0.926 & 0.029 & 0.038 & 0.934    \\
UAL~\cite{ual}           & 0.045 & 0.021 & 0.935 & 0.023 & 0.033 & 0.945    \\
DAEM~\cite{daem}         & \underline{0.042} & \underline{0.019} & \underline{0.940}  & \underline{0.021}  & \underline{0.031} & \underline{0.949}     \\ 
\midrule
Ours           & \textbf{0.038} & \textbf{0.016} & \textbf{0.946}  & \textbf{0.017}  & \textbf{0.028} & \textbf{0.955}  \\
\bottomrule[1.2pt]
\end{tabular}}
\caption{Quantitative comparison of different methods on two real datasets. The best results are in \textbf{bold}, while the second-best results are in \underline{underlined}.}
\label{table_real_compare}
\end{table}

\begin{table}
\centering

\resizebox{\linewidth}{!}{\begin{tabular}{c|c|c|c|c|c} 
\toprule[1.2pt]
Methods & \begin{tabular}[c]{@{}c@{}}FLOPs\\(G)\end{tabular} & \begin{tabular}[c]{@{}c@{}}Param\\(M)\end{tabular} & \begin{tabular}[c]{@{}c@{}}Training\\Time\\(min)\end{tabular} & \begin{tabular}[c]{@{}c@{}}Test\\Time\\(s)\end{tabular} & \begin{tabular}[c]{@{}c@{}}Test\\Iterations\end{tabular}  \\   
\midrule
MHFNet~\cite{mhfnet}  & 21226    & 129.4    & 1340               & 21.36        & \textbf{1}       \\
EDIP-Net~\cite{EDIP}  & 394.53    & 3.51    & /               & 1594.60        & 7000       \\
CUCaNet\cite{cucanet} & 7226     & 2.28     & /                  & 1806.18         & 10000            \\
UAL~\cite{ual}        & 3599     & 7.13     & 325                & 28.36        & 1500             \\
DAEM~\cite{daem}      & \textbf{1.67}     & \textbf{0.02}    & \textbf{119}                & \underline{19.56}        & \underline{250}              \\
Ours                  & \underline{279.42}   & \underline{0.55}    & \underline{215}                & \textbf{0.15}         & \textbf{1}                \\
\bottomrule[1.2pt]
\end{tabular}}
\caption{Efficiency comparison across methods, with best and second-best results in \textbf{bold} and \underline{underline}, respectively.}
% \caption{Efficiency comparison among different methods. The best results are in \textbf{bold}, while the second-best results are in \underline{underlined}.}
\vspace{-0.8cm}
\label{table_complex}
\end{table}

% \begin{figure}[!hp]
% 	\centering
% 	\includegraphics[scale=0.122]{figure/violin2.png}
% 	\caption{The distribution of two measurement metrics (PSNR, SAM) for the different methods under various degradations. The black lines and the white tangles denote medium and mean values.}
% 	\label{fig_violin}
% \end{figure}

% \begin{figure*}
% 	\centering
% 	\includegraphics[scale=0.255]{figure/ablation-3D.png}
% 	\caption{The reconstruction error comparison for the 8th band of the HR-HSIs from the CAVE dataset using different ablation experiments.
%     % The reconstruction error comparison of different ablation experiments.
%     }
% 	\label{fig_ablation_3D}
% \end{figure*}

\begin{figure}
	\centering
	\includegraphics[scale=0.463]{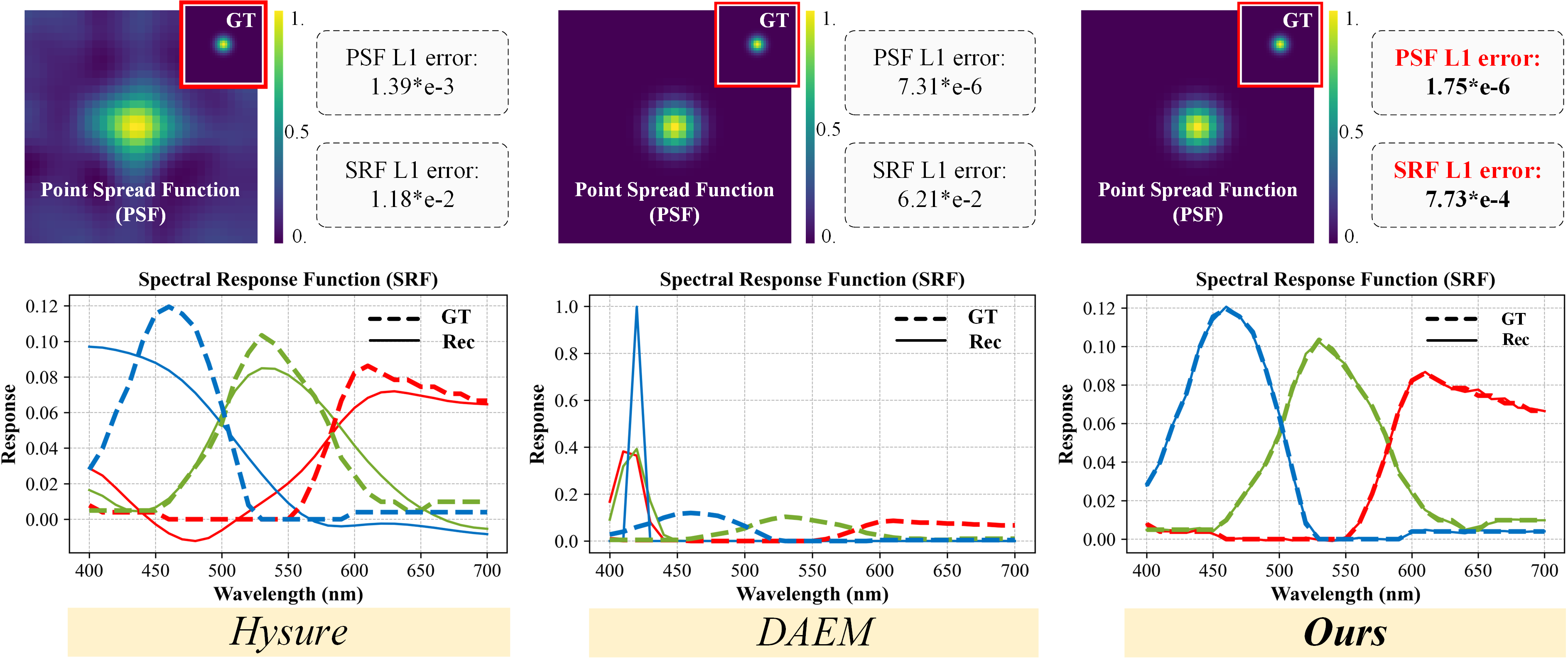}
	\caption{Comparison of degradation parameters PSF and SRF estimated by different methods.
	}
	\label{fig_degradation}
\end{figure}

\noindent \textbf{Real Dataset.} 
For real-world remote sensing datasets such as NCALM and WorldView-2, Table~\ref{table_real_compare} presents a comprehensive comparison between MossFuse and several state-of-the-art methods. Our approach achieves the best performance across all no‑reference metrics, clearly demonstrating its ability to deliver consistent, high‑quality results across diverse real‑world remote sensing scenarios.

To further assess its effectiveness, we conducted qualitative visual evaluations using synthetic pseudo‑RGB images generated from the reconstructed HSIs. As shown in Fig.~\ref{fig_NCALM}, the results confirm that MossFuse excels in preserving spatial details and maintaining spectral consistency compared to competing methods. In particular, our method delivers sharper edges in architectural elements, better preserves fine textures in complex urban structures, and produces more natural color rendering while effectively suppressing spectral artifacts.

Collectively, these quantitative and qualitative results highlight MossFuse's superior performance in real-world HSI-MSI fusion. By achieving an optimal balance of spatial detail enhancement and spectral fidelity without requiring ground-truth data, MossFuse proves to be a robust and reliable solution for practical remote sensing applications.

\begin{figure}[!hp]
	\centering
	\includegraphics[scale=0.09]{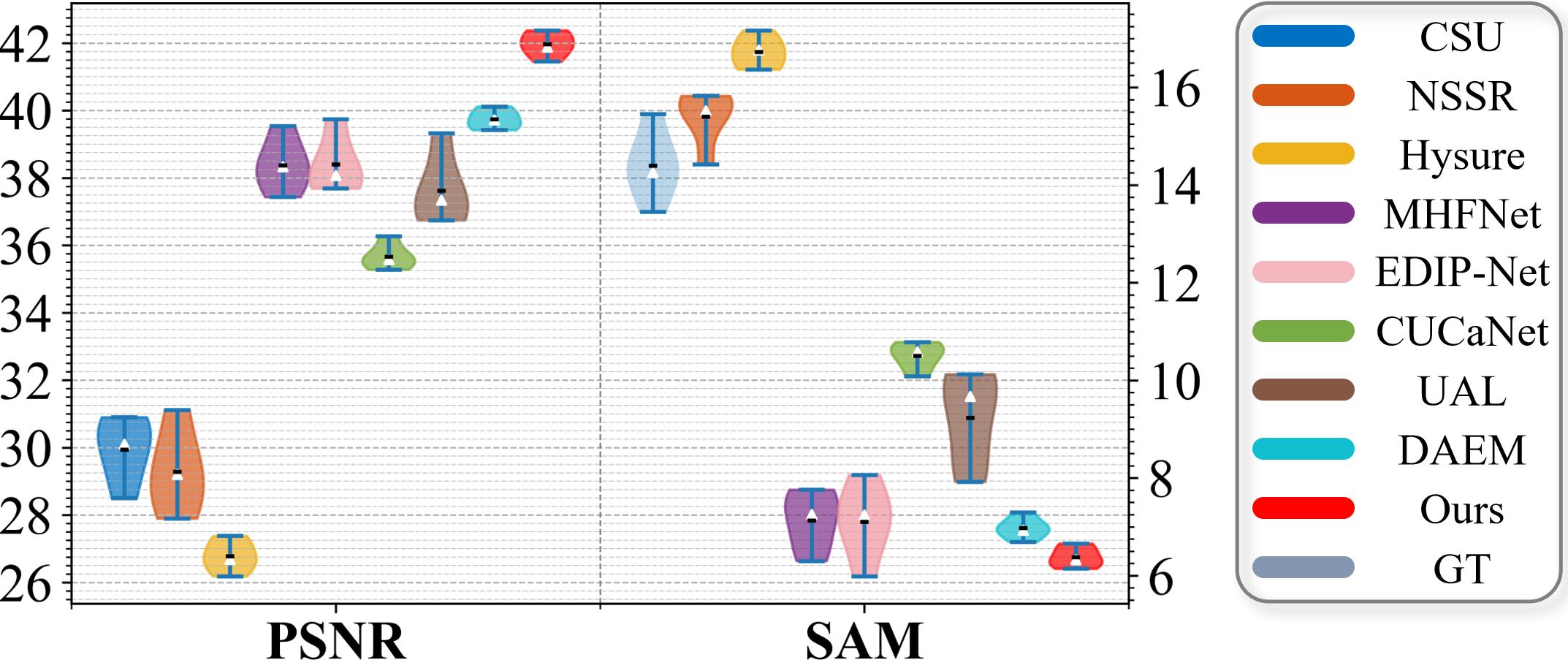}
	\caption{The distribution of two measurement metrics (PSNR, SAM) for the different methods under various degradations. The black lines and the white tangles denote medium and mean values.}
	\label{fig_violin}
\end{figure}

\subsection{Algorithmic Analysis}
\noindent \textbf{Computational Cost Analysis.} 
We present a comprehensive comparison of computational performance metrics in Table~\ref{table_complex}, evaluated on the CAVE dataset. Notably, methods like DAEM~\cite{daem} and UAL~\cite{ual} require sample-specific fine-tuning during inference, involving multiple optimization steps (see Table~\ref{table_complex}, last column), which substantially increases their actual runtime beyond what single-inference FLOPs suggest and limits their adaptability to dynamic scenarios.

In contrast, MossFuse achieves efficient inference by preserving only the essential components: multi-modality decoupling and modality aggregation processes, allowing high-quality HR-HSI reconstruction in a single-shot inference. This simple but effective architecture significantly enhances computational efficiency compared to methods requiring iterative optimization or fine-tuning during inference, making it suitable for real-world applications where processing time and resource constraints are critical factors.

% We present a comprehensive comparison of computational performance metrics in Table~\ref{table_complex}, evaluated on the CAVE dataset. It is worth mentioning that certain methods, such as DAEM~\cite{daem} and UAL~\cite{ual} incorporate a fine-tuning mechanism for each input sample during inference, requiring multiple optimization iterations (as specified in the last column of Table~\ref{table_complex}) to achieve optimal results. This iterative refinement paradigm substantially increases their actual computational overhead compared to single-inference approaches, explaining why their execution times exceed what would be expected from their single-inference FLOPs count alone while exposing fundamental generalizability constraints in adaptation to dynamic imaging conditions.

% In contrast, MossFuse maintains an efficient inference pipeline by retaining only the essential multi-modality decoupling and modality aggregation processes, enabling high-quality HR-HSI generation in a single-shot inference. This simple architecture significantly enhances computational efficiency compared to methods requiring iterative optimization or fine-tuning during inference, making it suitable for real-world applications where processing time and resource constraints are critical factors.

\noindent \textbf{Degradation Estimation Analysis.}
We conduct a comprehensive analysis of degradation parameter estimation by comparing our method with several fully blind approaches, including HySure~\cite{simoes2014convex} and DAEM~\cite{daem}, as illustrated in Fig.~\ref{fig_degradation}. Our algorithm demonstrates exceptional accuracy in estimating degradation parameters across both spatial and spectral dimensions, achieving near-optimal results.

Moreover, given the diversity of degradation models in real-world scenarios, we further assess the framework’s robustness by evaluating the performance distribution of different algorithms under six groups of degradation conditions~\cite{daem} on the CAVE dataset. Fig.~\ref{fig_violin} presents violin plots of two evaluation metrics, PSNR and SAM, illustrating the reconstruction performance across various degradation settings. These results demonstrate that our algorithm excels not only in accuracy but also in robustness compared to others.

In unsupervised hyperspectral and multispectral image fusion (HMIF) tasks, the absence of ground truth implies that the accuracy of degradation model estimation is crucial for high-fidelity HR-HSI reconstruction. Given the remarkable results achieved, a critical question naturally arises: Is the observed state-of-the-art performance primarily attributed to the degradation estimation module, rather than the core modality decoupling architecture?

To investigate this aspect, we assume knowledge of the degradation model and adapt several state-of-the-art (SOTA) fully supervised fusion methods to operate under the same unsupervised setting for fair comparison. 

In this setup, the loss function is constructed with both the HR-MSI and the LR-HSI as dual reconstruction constraints, rather than their original supervised losses. Formally, the loss is defined as:
\begin{equation}
    \mathcal{L} = \|\text{Y} - \hat{\text{X}}R \|_1 + \| x - C\hat{\text{X}} \|_1,
\end{equation}
where \( \|\cdot\| \) represents the L1 norm, \(\hat{\text{X}}\) is the reconstructed HR-HSI, and the degradation operators \(R\) (spectral downsampling) and \(C\) (spatial downsampling) are fixed and given. Such a self-supervised formulation enables effective model training without requiring access to ground-truth HR-HSI, making it suitable for the unsupervised fusion task.

The adapted methods span a diverse range of recent architectural paradigms: (i) Transformer-based models: 3DT-Net~\cite{ma2023learning} and DCTransformer~\cite{ma2024reciprocal} employ transformer blocks to capture long-range spatial–spectral dependencies via self-attention, thus overcoming CNNs’ limited receptive field and enhancing feature representation for hyperspectral reconstruction; (ii) Mamba-based models: SRLF-Net~\cite{liu2025selective} adopts a selective re-learning mechanism to refine degraded features, while SINet~\cite{sinet} fuses Mamba’s long-range modeling with CNN’s local extraction to adaptively enhance global–local complementarity; and (iii) Frequency-domain-based models: FSDFF~\cite{tan2024frequency} learns frequency correlations to improve spectral fidelity, and FeINFN~\cite{liang2024fourier} jointly models spatial and Fourier domains to enhance high-frequency details and expand the receptive field. 

For fair comparison, our MossFuse is trained under the same SRF and PSF degradations. As shown in Table~\ref{table_ablation_degradation}, MossFuse surpasses all supervised baselines on the CAVE dataset (PSNR, SSIM, SAM, ERGAS), verifying its superior reconstruction accuracy in an unsupervised setting.

\begin{table}
\centering
\resizebox{\linewidth}{!}{\begin{tabular}{c|c|c|c|c|c} 
\toprule[1.2pt]
\rowcolor{Gray1} Methods                 & Deg. Model & PSNR $\uparrow$  & SSIM $\uparrow$  & SAM $\downarrow$   & ERGAS $\downarrow$ \\ 
\midrule
\multirow{2}{*}{MHFNet~\cite{mhfnet} }  & \textit{unknown}   & 38.45    & 0.970    & 6.88     & 1.87  \\
                                        & \textit{known}     & 39.96    & 0.978    & 6.32     & 1.06  \\
\midrule
\multirow{2}{*}{EDIP-Net~\cite{EDIP} }  & \textit{unknown}   & 38.77    & 0.977    & 6.81     & 1.01  \\
                                        & \textit{known}     & 40.16    & 0.980    & 6.60     & 0.86  \\
\midrule
\multirow{2}{*}{CUCaNet\cite{cucanet}}  & \textit{unknown}   & 35.46    & 0.929    & 10.43    & 0.97   \\
                                        & \textit{known}     & 40.06    & 0.980    & 6.83    & 0.53   \\ 
\midrule
\multirow{2}{*}{UAL~\cite{ual}}         & \textit{unknown}   & 37.78    & 0.976    & 9.24     & 0.68  \\
                                        & \textit{known}     & 39.86    & 0.980    & 7.24     & 0.56 \\ 
\midrule
\multirow{2}{*}{DAEM~\cite{daem}}   & \textit{unknown}  & \underline{39.83} & \underline{0.978} & \underline{6.79}  & \underline{0.54} \\ 
                                    & \textit{known}    & 40.98 & 0.983 & 6.60 & 0.48 \\
\midrule
\rowcolor{blue!10} 3DT-Net~\cite{ma2023learning}   & \textit{known}    & 40.52 & 0.980 & 6.67 & 0.52 \\
\midrule
\rowcolor{blue!10} DCTransformer~\cite{ma2024reciprocal}   & \textit{known}    & 42.34 & \underline{0.984} & 6.52 & \underline{0.42} \\
\midrule
\rowcolor{blue!10} SRLF-Net~\cite{liu2025selective}   & \textit{known}    & \underline{42.56} & 0.983 & \underline{6.51} & 0.44 \\
\midrule
\rowcolor{blue!10} SINet~\cite{sinet}   & \textit{known} & 41.92 & 0.979 & 6.57 & 0.45 \\
\midrule
\rowcolor{blue!10} FSDFF~\cite{tan2024frequency}   & \textit{known}    & 41.86 & 0.981 & 6.57 & 0.48 \\
\midrule
\rowcolor{blue!10} FeINFN~\cite{liang2024fourier}   & \textit{known}    & 41.68 & 0.982 & 6.60 & 0.49 \\
\midrule
\multirow{2}{*}{Ours}           & \textit{unknown}   & \textbf{42.15} & \textbf{0.983} & \textbf{6.48}  & \textbf{0.42} \\
                                & \textit{known}     & \textbf{42.86} & \textbf{0.986} & \textbf{6.43}  & \textbf{0.40} \\
\bottomrule[1.2pt]
\end{tabular}}
\caption{Quantitative comparison of various methods on CAVE dataset with known and unknown degradation models with best and second-best results in \textbf{bold} and \underline{underlined}, respectively.}
\vspace{-0.6cm}
\label{table_ablation_degradation}
\end{table}

\begin{figure}[!hp]
	\centering
	\includegraphics[width=0.9\linewidth]{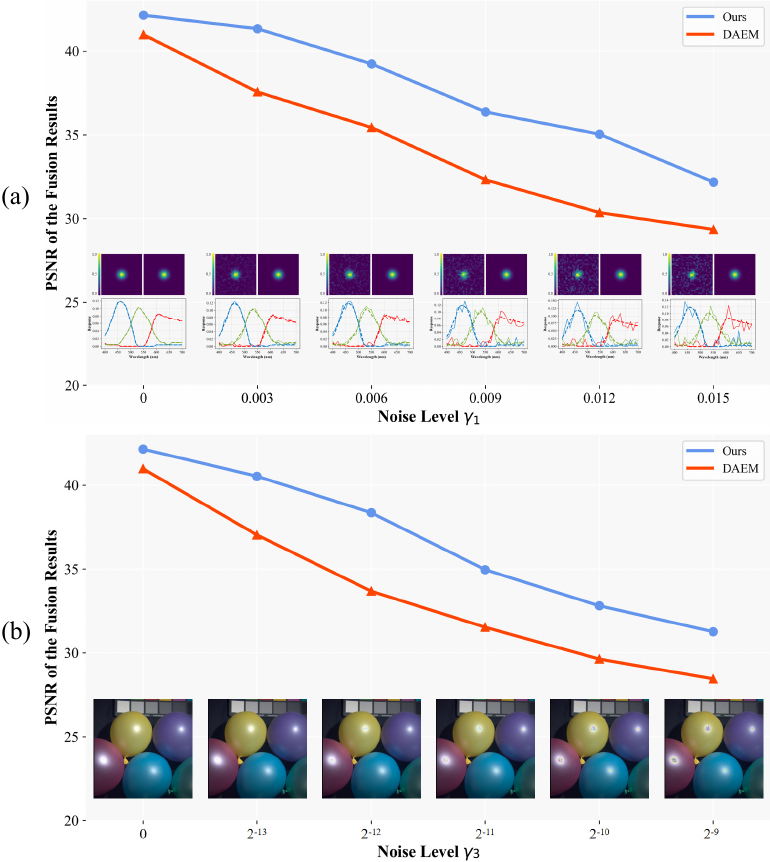}
	\caption{Performance under mismatched degradation conditions. (a) Degradation with Gaussian noise. The bottom row visualizes the corresponding corrupted degradation functions. For PSF, the left side shows the corrupted version(\(C'\)), while the right side shows the ideal (clean) PSF(\(C\)). For SRF, the solid line represents the corrupted SRF(\(R'\)), whereas the dashed line indicates the ideal SRF(\(R\)). (b) Degradation with Poisson noise. The bottom row visualizes input example corrupted by Poisson noise.}
	\label{fig_degradation_noise}
\end{figure}

With access to precise degradation parameters, all unsupervised methods exhibit notable performance improvements, with CUCaNet showing the most significant gains. The two supervised methods yield even better spectral accuracy, likely due to their larger parameter capacity. Nevertheless, our approach consistently outperforms all baselines across four evaluation metrics, affirming the effectiveness and generalizability of the proposed modality decomposition paradigm in unsupervised HMIF tasks.

To further evaluate the robustness and generalization capability of our method under real-world conditions, we investigate how performance is affected when the estimated degradation parameters deviate from the ideal or ground-truth settings. Such deviations can arise from two sources: (1) the inherent limitations of the network in accurately estimating degradation parameters, and (2) real-world deviations from ideal assumptions due to sensor noise, atmospheric interference, system calibration errors, and other factors.

Based on the CAVE dataset, we first simulate these effects by injecting controlled Gaussian noise into both the SRF(\(R\)) and PSF(\(C\)) during the inference phase. This procedure generates HR-MSI and LR-HSI inputs that differ from the training distribution, thereby introducing a domain shift and simulating practical scenarios involving inaccurate or mismatched degradation estimation.

Specifically, given the degradation parameters \(R\) and \(C\) used during training, we introduce two perturbation control factors, \(\gamma_1\) and \(\gamma_2\), to regulate the intensity of additive Gaussian noise applied to SRF and PSF, respectively:
\begin{equation}
    R' = R + \mathcal{N}(0, \gamma_1^{2}), \quad C' = C + \mathcal{N}(0, \gamma_2^{2}).
\end{equation}

We set \(\gamma_1 \in\{0, 3*10^{-3}, 6*10^{-3}, 9*10^{-3}, 12*10^{-3}, 15*10^{-3} \} \), and \(\gamma_2 = \gamma_1 / 3\} \), reflecting the fact that SRF and PSF exhibit different sensitivities to noise. 

These perturbed degradation parameters \(R'\) and \(C'\) are then used to synthesize test samples, ensuring a domain shift between training and testing. The results are illustrated in Fig.~\ref{fig_degradation_noise} (a). We present the PSNR values of our method (Ours) and a representative baseline (DAEM) under varying noise intensities. The bottom part of the figure displays the corrupted PSF and SRF corresponding to each noise level. 

Beyond Gaussian noise, Poisson noise is incorporated to simulate degradation estimation errors arising from inherent sensor noise. Since Poisson noise is signal-dependent, we simulate it by first scaling the normalized input image \(\text{Y},x\in [0,1]\) by a gain factor \(\delta\), then sampling from a Poisson distribution with mean \(\delta \cdot \text{Y}, \delta \cdot x\), yielding \(\text{Y}_{\text{noisy}} \sim \mathcal{P}(\delta \cdot \text{Y})\) and \(x_{\text{noisy}} \sim \mathcal{P}(\delta \cdot x)\) respectively, and finally rescaling the result back to \([0,1]\). We set \(\gamma_3=1/\delta\), with \(\gamma_3 \in\{0, 2^{-13}, 2^{-12}, 2^{-11}, 2^{-10}, 2^{-9} \} \), corresponding to varying noise levels. Smaller values of \(\gamma_3\) amplify the relative noise intensity, simulating low-light or high-noise acquisition conditions.

As shown in Fig.~\ref{fig_degradation_noise}, our method achieves consistently high performance under both Gaussian and Poisson noise conditions. Even in the presence of severe degradation, MossFuse significantly outperforms DAEM, demonstrating strong robustness across diverse and realistic noise environments.

These experiments highlight the robustness of our framework under imperfect degradation estimation and confirm its suitability for real-world deployment.

\begin{figure}
	\centering
	\includegraphics[scale=0.5]{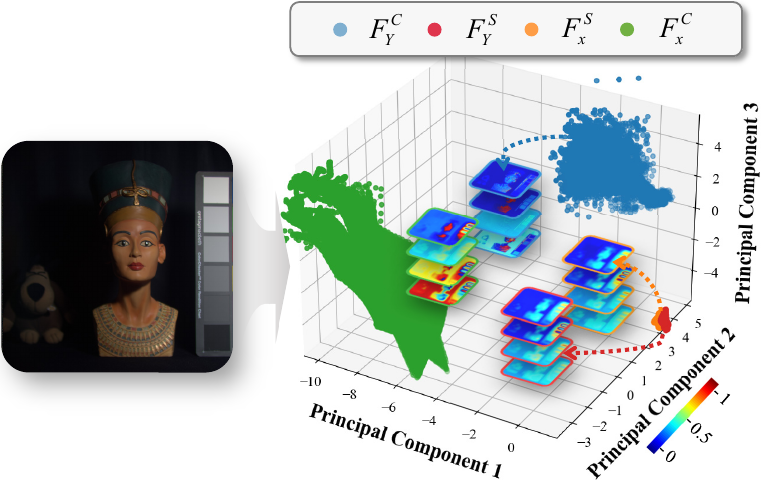}
	\caption{Four decoupled features visualization and distribution after PCA, showing the separation between modality-shared and modality-complementary representations.}
	\label{fig_pca}
\end{figure}

\begin{figure*}
    \centering
    \begin{subfigure}{0.31\textwidth}
        \includegraphics[width=\linewidth]{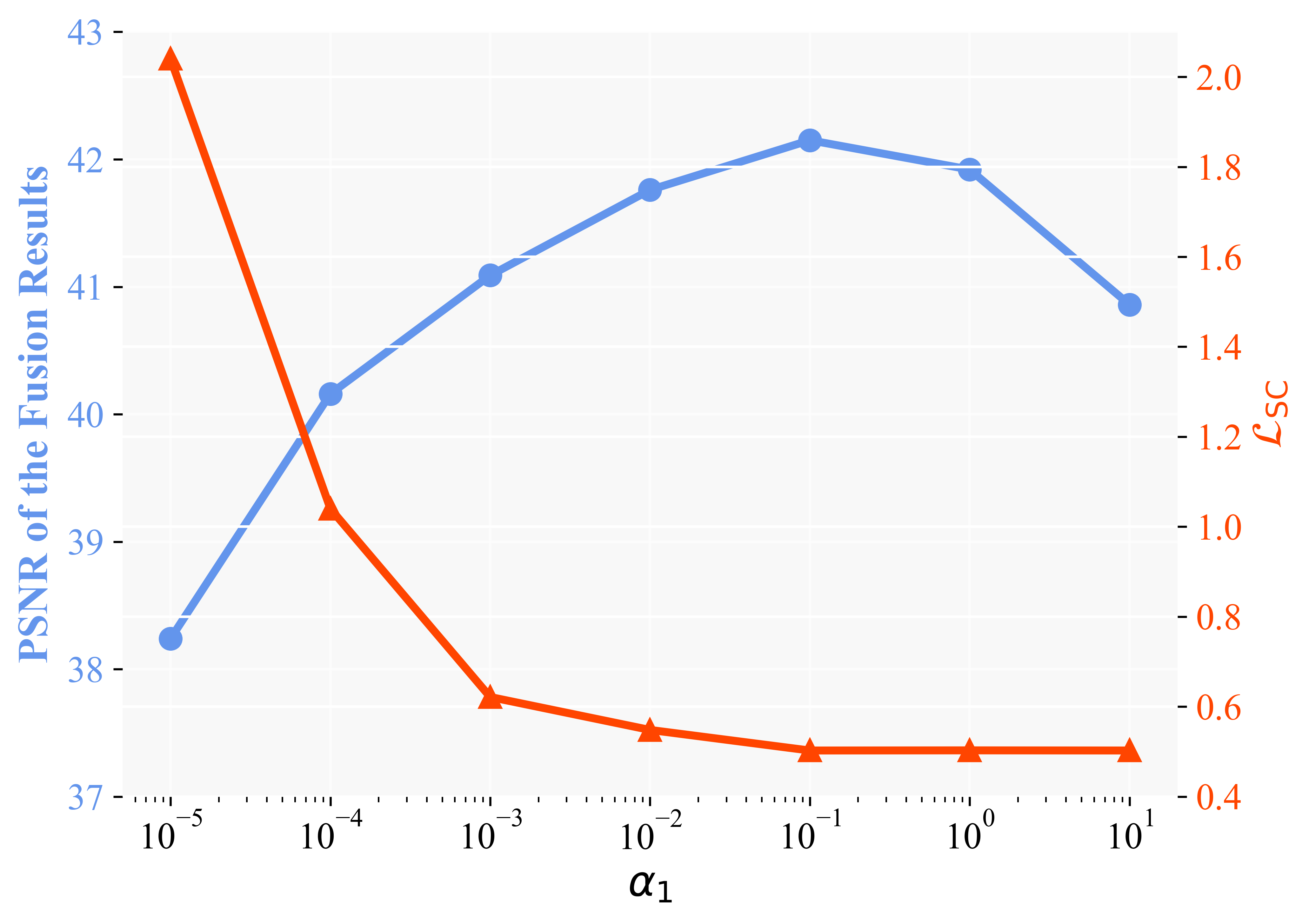}
        \caption{Performance with different \(\alpha_1\)}
        \label{fig:alpha1}
    \end{subfigure}%
    \hfill
    \begin{subfigure}{0.31\textwidth}
        \includegraphics[width=\linewidth]{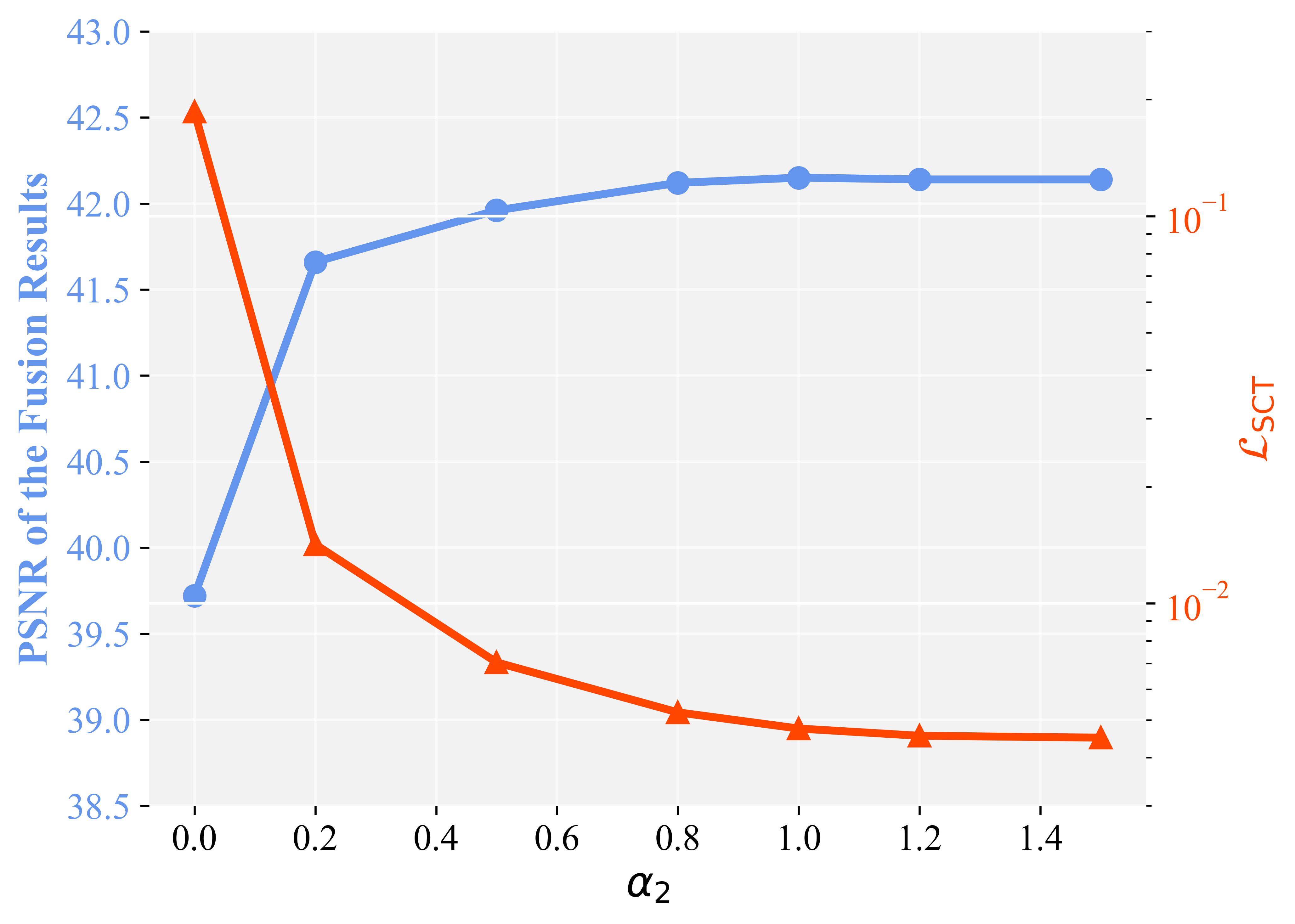}
        \caption{Performance with different \(\alpha_2\)}
        \label{fig:alpha2}
    \end{subfigure}%
    \hfill
    \begin{subfigure}{0.31\textwidth}
        \includegraphics[width=\linewidth]{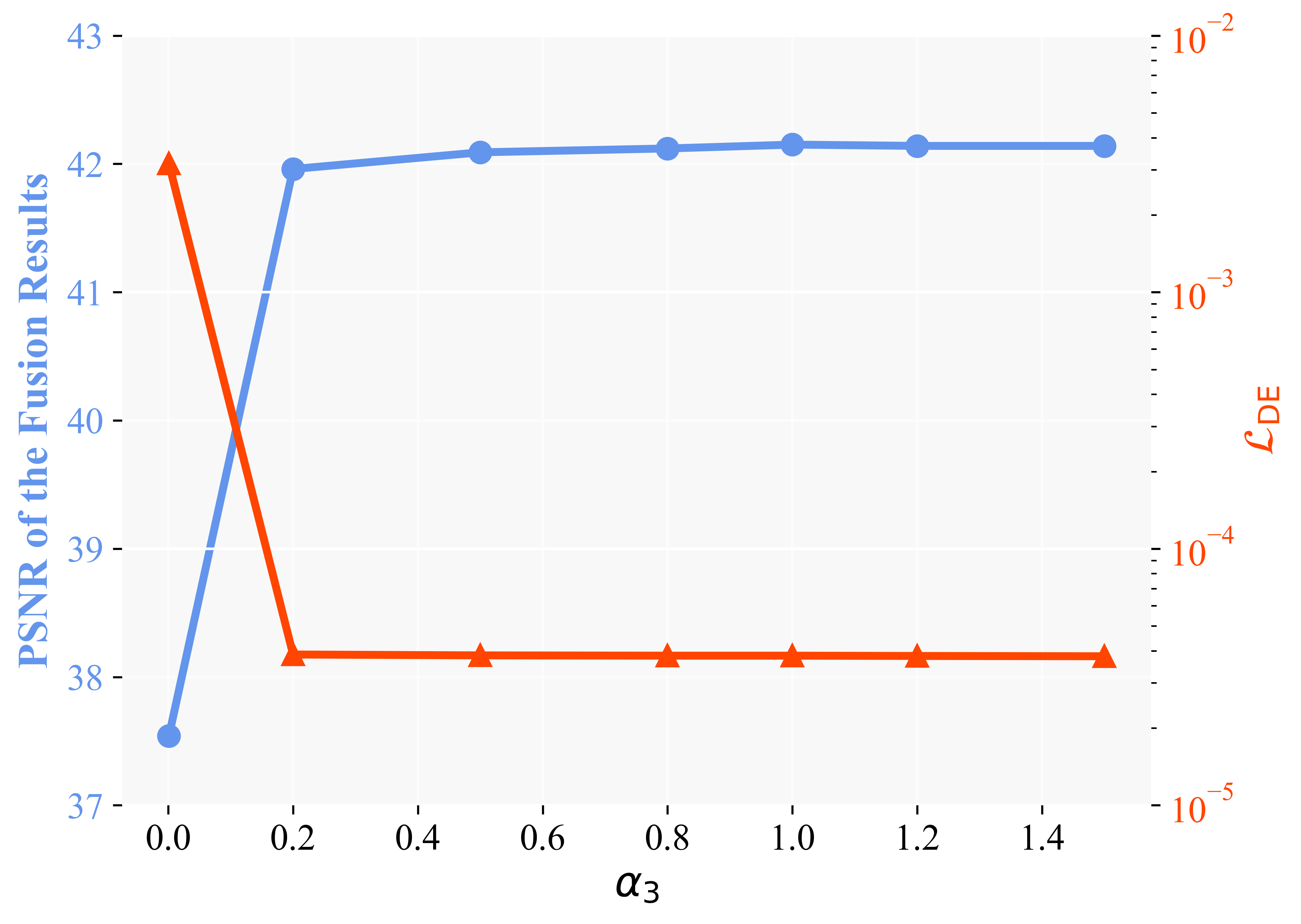}
        \caption{Performance with different \(\alpha_3\)}
        \label{fig:alpha3}
    \end{subfigure}
    \caption{Comparison of algorithm's performance with different settings of the training loss weight parameters \(\alpha_1\), \(\alpha_2\), and \(\alpha_3\).}
    \label{fig_alpha}
\end{figure*}

\noindent \textbf{Design Paradigm Analysis.} 
Our framework explicitly separates shared and complementary modalities, which adopts a design choice that fundamentally differs from both classical and recent approaches. Traditional linear decomposition methods (e.g., CNMF~\cite{yokoya2011coupled}, Tucker~\cite{kanatsoulis2018hyperspectral}) approximate hyperspectral–multispectral fusion under linear mixing or low-rank priors, which, while interpretable, cannot fully capture nonlinear modality interactions. 

Model-driven tensor decomposition methods such as deep low-rank tensor~\cite{dian2024spectral}, Generalized Tensor Nuclear Norm (GTNN)~\cite{dian2024hyperspectral}, and tensor-ring/subspace mechanism~\cite{xu2021hyperspectral} leverage structured priors (e.g., low-rankness, factorization) to capture high-order correlations. However, they typically rely on strong assumptions (e.g., fixed ranks, tensor formats), require careful hyperparameter tuning, and are sensitive to model specification, limiting their flexibility and generalization.

Recent learning-based frameworks such as DAEM~\cite{daem} and UAL~\cite{ual} take different routes: DAEM emphasizes iterative degradation modeling and multi-stage decomposition, while UAL relies on unsupervised adaptation and test-time optimization, updating parameters per sample. 

However, all these approaches lack an explicit mechanism for structural disentanglement of shared and complementary information. In contrast, MossFuse enforces modality-specific subspaces through subspace clustering and self-supervised reconstruction, enabling efficient, single-shot fusion without per-sample optimization.

\noindent \textbf{Modality Decoupling Capability Analysis.}
Our methodology fundamentally decouples the HR-MSI and LR-HSI into two distinct components: modality-shared components \( F^{S}_{\text{Y}} \) and \( F^{S}_{x} \), represented by the latent LR-MSI, and modality-complementary components \( F^{C}_{\text{Y}} \) and \( F^{C}_{x} \), consisting of spatial and spectral complementary information, respectively. To demonstrate the effectiveness of this decoupling process, we first visualize the above four critical decoupled features and further project them into a low-dimensional space using Principal Component Analysis (PCA), as shown in Fig.~\ref{fig_pca}. 

The visualization and PCA results reveal two key observations: (1) The two modality-shared features \( F^{S}_{\text{Y}} \) and \( F^{S}_{x} \) have nearly identical visual patterns and show highly overlapping distributions in the PCA space, indicating a strong consistency in the decoupling process. Since LR-MSI contains limited redundant information, the extracted shared features tend to form compact and concentrated clusters, reflecting their aligned representation across modalities; (2) The modality-complementary features demonstrate distinct characteristics. The spatial complementary features (\( F^{C}_{\text{Y}} \)) contain rich spatial textures with relatively low spectral variation, while the spectral complementary features (\( F^{C}_{x} \)) emphasize spectral diversity with less spatial structure. These two types of complementary feature are separated from each other as well as from the shared features within the PCA space, highlighting their individual roles: each of which encodes complex residual information specific to their modality. This naturally results in more dispersed and modality-specific feature distributions.

% The spatial complementary features exhibit richer spatial details and less variation in channel dimensions, while the spectral complementary features show the opposite. Moreover, their distribution form distinct clusters, clearly separated from both shared and each other’s representation. This divergence reflects their specialized roles: spatial (\( F^{C}_{\text{Y}} \)) and spectral (\( F^{C}_{x} \)) components necessitate distributed encoding of complex residual information, resulting in a more dispersed PCA distribution. 

These findings validate MossFuse’s effectiveness in disentangling and preserving both shared and modality-specific representations.

\noindent \textbf{Hyper-parameter Analysis.}
\label{parameter}
Recall that our training loss is formulated as:
\begin{equation}
\mathcal{L}_{\text{total}} = \mathcal{L}_{\text{MA}}+\alpha_1\mathcal{L}_{\text{SC}}+\alpha_2\mathcal{L}_{\text{SCT}}+\alpha_3\mathcal{L}_{\text{DE}},
\end{equation}
where $\mathcal{L}_{\text{MA}}$ denotes the modality aggregation loss (supervises the final reconstruction results), $\mathcal{L}_{\text{SC}}$ is the subspace clustering loss,  $\mathcal{L}_{\text{SCT}}$ is the self-supervised constraint loss, and $\mathcal{L}_{\text{DE}}$ corresponds to the degradation estimation loss. Each loss term is designed to address a specific aspect of the fusion task, and their contributions are balanced by the corresponding weights \(\alpha_1\), \(\alpha_2\), and \(\alpha_3\). 

To comprehensively evaluate the sensitivity of the proposed framework to these hyperparameters and to further understand the contribution of each loss component, we perform a series of experiments by individually varying one hyperparameter while keeping the others fixed. The experimental results, summarized in Fig.~\ref{fig_alpha}, illustrate how different configurations of loss weighting influence the final fusion performance (PSNR, \textcolor{Blue1}{blue curves}) and the corresponding loss values (\textcolor{Red2}{red curves}).

\textbf{1. Effect of varying \(\mathbf{\alpha_1}\).} 
Fig.~\ref{fig_alpha} (a) shows the relationship between the PSNR metric and the subspace clustering loss \(\mathcal{L}_{\text{SC}}\) under varying values of \(\alpha_1 \in \{10^{-5},10^{-4},10^{-3},10^{-2},10^{-1},10^{0}, 10^{1}\}\). As \(\alpha_1\) increases, \(\mathcal{L}_{\text{SC}}\) consistently decreases and eventually stabilizes around 0.5. The PSNR, however, exhibits a bell-shaped curve: it first increases and then decreases, reaching a peak near \(\alpha_1=0.1\). This observation can be interpreted as follows: when \(\alpha_1\) is small, the \(\mathcal{L}_{\text{SC}}\) is insufficiently weighted, rendering it ineffective in guiding the accurate decomposition of modality-shared features (\( F^{S}_{\text{Y}} \) and \( F^{S}_{x} \)), and modality-complementary features (\( F^{C}_{\text{Y}} \), and \( F^{C}_{x} \)). Conversely, when \(\alpha_1=0.1\) becomes too large, the overemphasis on \(\mathcal{L}_{\text{SC}}\) disrupts the optimization balance among different modules, which in turn degrades the overall fusion performance and leads to a drop in PSNR.

\textbf{2. Effect of varying \(\mathbf{\alpha_2}\).} 
The self-supervised constraint loss $\mathcal{L}_{\text{SCT}}$ is designed to preserve the integrity and fidelity of the decoupled modality features. While its effectiveness has been preliminarily verified through ablation (by removing it entirely and partially), we further investigate its influence by varying \(\alpha_2 = \{10^{-4},0.2,0.5,0.8,1,1.2,1.5\}\), as shown in Fig.~\ref{fig_alpha} (b). When \(\alpha_2\) is too small, \(\mathcal{L}_{\text{SCT}}\) contributes minimally, preventing the model from fully leveraging the self-supervised constraints, which results in weak feature regularization, suboptimal convergence, and degraded PSNR performance. As \(\alpha_2\) increases, both the loss and the final PSNR gradually stabilize, indicating that moderate values (e.g., around 1.0) provide a good balance between constraint enforcement and overall optimization.

\textbf{3. Effect of varying \(\mathbf{\alpha_3}\).} 
The degradation estimation loss \(\mathcal{L}_{\text{DE}}\) is leveraged to learn the spatial and spectral degradation parameters and further mitigate the ill-posed nature of unsupervised HMIF. To quantify its impact, we varied \(\alpha_3 \in \{10^{-4},0.2,0.5,0.8,1,1.2,1.5\}\), and recorded the corresponding changes in PSNR and \(\mathcal{L}_{\text{DE}}\), as shown in Fig.~\ref{fig_alpha} (c). 
The results reveal that when \(\alpha_3\) is too small, the model struggles to effectively learn the degradation patterns, resulting in poor convergence and fusion quality. Once \(\alpha_3\) exceeds a certain threshold (e.g., 0.2), \(\mathcal{L}_{\text{DE}}\) converges, and further weight increases have a negligible impact on performance, indicating a moderate weighting of \(\mathcal{L}_{\text{DE}}\)  is sufficient to achieve optimal results.

\noindent \textbf{Downstream Experiments Analysis.} Beyond reconstruction fidelity, the value of HR-HSI lies in its utility for downstream vision tasks. We evaluate MossFuse on representative applications including classification and material identification to validate its practical effectiveness as below.

\textbf{1. Classification.}: To evaluate the benefit of fusion for semantic interpretation, we conduct a classification experiment on the Houston dataset (349×1905 spatial, 144 spectral bands), using SpectralFormer~\cite{hong2021spectralformer} as the classifier under standard settings. We simulate LR-HSI using applying an 8×8 Gaussian blur (\(\sigma\)=2) followed by ×4 spatial downsampling, and generate HR-MSI by averaging every 36 adjacent spectral bands. The fused HR-HSI is then input to the classifier and compared against the upsampled LR-HSI (without fusion).

\begin{figure*}[htb]
	\centering
	\includegraphics[width=0.96\linewidth]{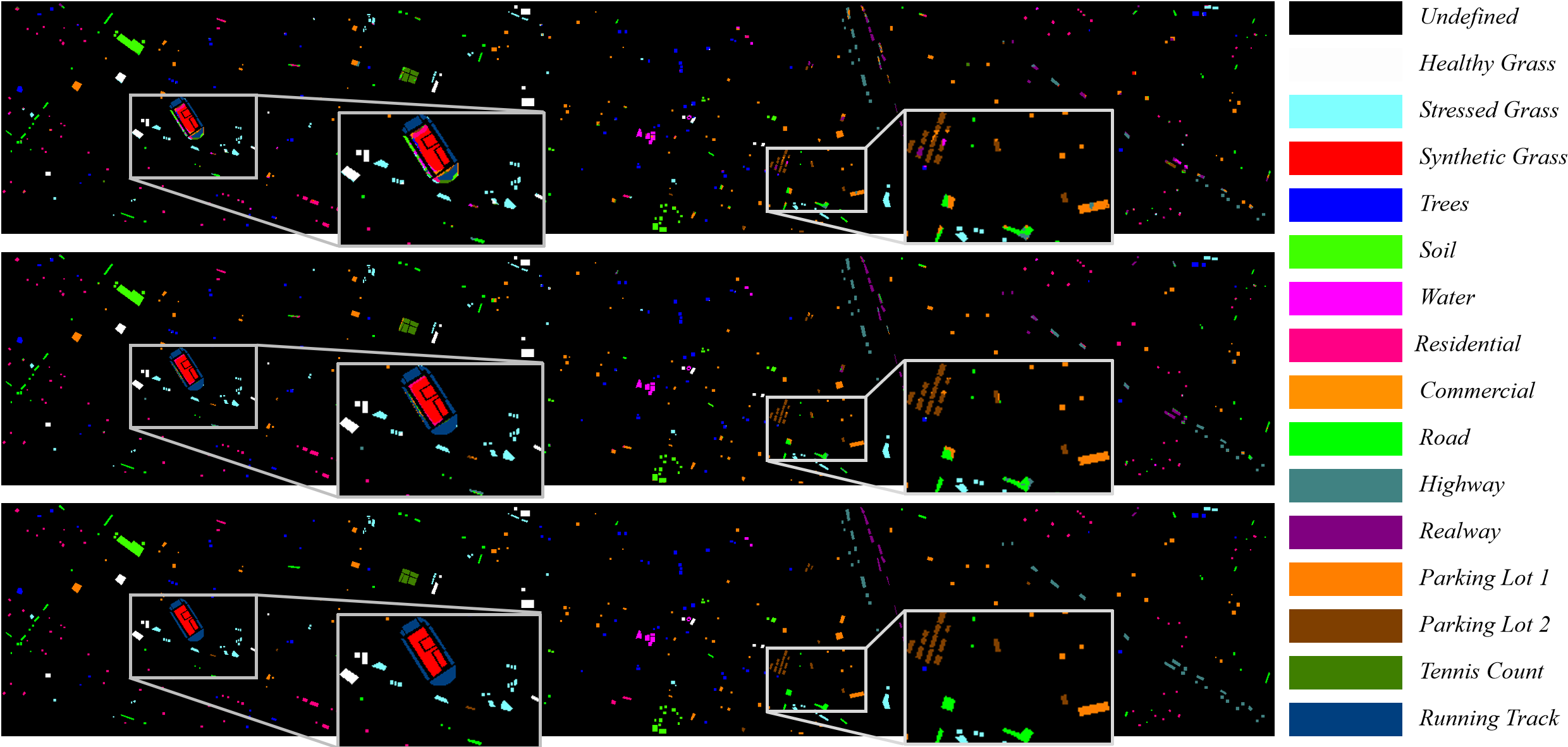}
    \caption{Classification results of LR-HSI (top), predicted HR-HSI (middle), and reference (bottom).}
    \label{fig_classification}
\end{figure*}

As shown in Table~\ref{table_classification}, the fused HR-HSI achieves higher accuracies across most categories. The visual comparison provided Fig.~\ref{fig_classification} further highlights regions where MossFuse enables more accurate classification, particularly in spectrally ambiguous or fine-structured areas. This demonstrates that our method preserves discriminative spectral information beneficial for semantic understanding.

\begin{figure}[htb]
	\centering
	\includegraphics[width=0.9\linewidth]{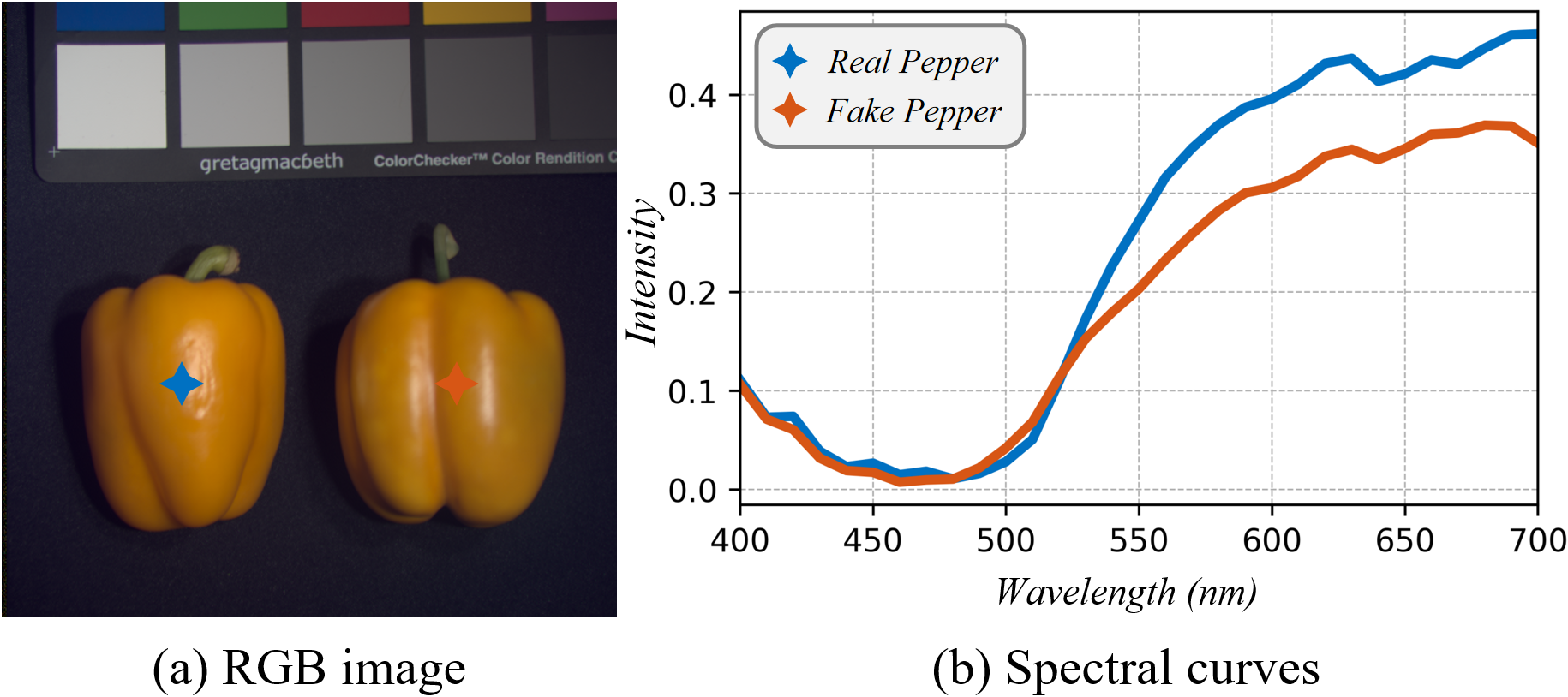}
     \caption{Application for material discrimination.}
     \label{fig_material}
\end{figure}

\textbf{2. Material identification.} To assess the fidelity of reconstructed spectra for physical interpretation, we examine material distinguish ability using real and fake peppers from the CAVE dataset (Fig.~\ref{fig_material}). We extract and compare their spectral signatures from the fused HR-HSI. As shown, the two materials exhibit readily distinguishable reflectance curves, especially beyond 500nm, despite similar visual appearance. This confirms that MossFuse recovers physically meaningful spectral variations, enabling reliable material differentiation.

\begin{table}
\centering
\resizebox{\linewidth}{!}{\begin{tabular}{cccc} 
\toprule[1.2pt]
\multirow{2}{*}{Category}   &\multicolumn{2}{c}{Accuracy(\%)} \\
\cmidrule(r){2-3} 
& LR-HSI   & Predicted HR-HSI  &\\ 
\midrule
Healthy grass    & \textbf{85.29}  & 83.29 & \\
Stressed grass   & 97.05  & \textbf{99.06} & \\
Synthetic grass  & 74.75  & \textbf{83.56} & \\
Trees            & 93.01  & \textbf{95.17} & \\
Soil             & 96.05  & \textbf{99.24} & \\
Water            & 91.38  & \textbf{93.01} & \\
Residential      & 84.46  & \textbf{86.29} & \\
Commercial       & 70.58  & \textbf{71.98} & \\
Road             & \textbf{81.07}  & 78.47 & \\
Highway          & 58.03  & \textbf{66.51} & \\
Railway          & 66.72  & \textbf{74.67} & \\
Parking lot 1    & 66.50  & \textbf{82.52} & \\
Parking lot 2    & \textbf{67.80}  & 67.72 & \\
Tennis court     & 89.25  & \textbf{89.88} & \\
Running track    & 82.12  & \textbf{90.49} & \\
\midrule
Average accuracy & 80.27  & \textbf{84.12} & \\
\bottomrule[1.2pt]
\end{tabular}}
\caption{Classification results of LR-HSI and predicted HSI with best results in \textbf{bold}.}
\vspace{-0.6cm}
\label{table_classification}
\end{table}

% \begin{figure}
% 	\centering
% 	% \includegraphics[scale=0.54]{figure/PCA.pdf}
% 	\includegraphics[scale=0.46]{figure/PCA_l.pdf}
% 	\caption{Data distribution of four decoupled features after PCA, showing the separation between modality-shared and modality-complementary representations.}
% 	\label{fig_pca}
% \end{figure}

\begin{table*}
\centering
\resizebox{\linewidth}{!}{\begin{tabular}{c|c|cc|cccc|cccc} 
\toprule[1.2pt]
                    & \multirow{2}{*}{Configurations}  & \multirow{2}{*}{Parameters(M)$\downarrow$} & \multirow{2}{*}{Test Time(s)$\downarrow$}  & \multicolumn{4}{c}{CAVE~\cite{yasuma2010generalized}} & \multicolumn{4}{c}{Harvard~\cite{chakrabarti2011statistics}}
                   \\ \cmidrule(r){5-8} \cmidrule(r){9-12}
                   & & & & PSNR $\uparrow$  & SSIM $\uparrow$  & SAM $\downarrow$   & ERGAS $\downarrow$ & PSNR $\uparrow$  & SSIM $\uparrow$  & SAM $\downarrow$   & ERGAS $\downarrow$\\ 
\midrule
\multirow{3}{*}{1}  & w/o $\mathcal{L}_\text{SC}$  & 0.549  & 0.154 & 37.83  & 0.967  & 9.86  & 0.70  & 38.95 & 0.932 & 6.25  & 0.63 \\
                    & w/o aggre        & 0.549  & 0.154 & 39.86          & 0.975    & 7.88    & 0.56  & 40.45 & 0.942 & 5.63  & 0.55 \\
                    & w/o repel       & 0.549  & 0.154 & 38.74          & 0.972     & 8.46    & 0.61  & 40.03 & 0.938 & 5.82  & 0.60 \\ 
\midrule              
\multirow{3}{*}{2}  & res\_block   & 0.548  & 0.155 & 39.36     & 0.974     & 8.03  & 0.59            & 40.36 & 0.941 & 5.66  & 0.57 \\
                    & dilated conv  & \underline{0.547}  & \underline{0.153}  & 40.06   & 0.976     & 7.54   & 0.56           & 40.56 & 0.945 & 5.59  & 0.58 \\
                    & SWT        & 0.552  & 0.156  & 38.76      & 0.970    & 8.76    & 0.62           & 39.88 & 0.933 & 5.68  & 0.67 \\    
\midrule
\multirow{3}{*}{3}  & concat $\&$ conv & 0.551  & 0.156   & 38.92       & 0.970      & 8.93   & 0.61  & 40.02 & 0.936 & 5.62  & 0.59 \\
                    & concat $\&$ SWT  & 0.554  & 0.159   & 39.02       & 0.973      & 8.76  & 0.60   & 39.86 & 0.940 & 5.61  & 0.62 \\
                    & cross   & 0.552  & 0.156  & 40.29    & 0.976   & 7.42     & 0.52                & 41.88 & 0.950 & 4.94  & 0.54 \\ 
\midrule
\multirow{2}{*}{4}  & weighted fusion  & 0.553  & 0.154   & \underline{41.83}       & \underline{0.978}      & \underline{6.72}      & 0.46   & \underline{43.57} & \underline{0.988} & 3.84  & 0.42\\
% & 0.549  & 0.154  & 41.36    & 0.973   & 7.02     & 0.50  & 43.36 & 0.986 & 3.88  & 0.43 \\
%                     & concat $\&$ conv$^{2}$  & 0.550  & 0.155   & \underline{41.83}       & \underline{0.978}      & \underline{6.72}      & 0.46   & \underline{43.57} & \underline{0.988} & 3.84  & 0.42\\
                    & cross$^{2}$    & 0.552  & 0.156 & 41.29    & 0.971   & 7.06     & 0.51 & 43.39 & 0.985 & \underline{3.83}  & \underline{0.42} \\ 
\midrule
\multirow{2}{*}{5}  & w/o $\mathcal{L}_\text{SCT}$  & 0.549  & 0.154  & 39.72          & 0.973         & 7.75       & 0.53    & 41.87 & 0.956 & 4.84  & 0.52        \\
                    & w/o $\mathcal{L}_\text{SCT2}$ & 0.549  & 0.154  & 40.33         & 0.976          & 7.63       & \underline{0.45}  & 41.96 & 0.952 & 4.88  & 0.53 \\ 
\midrule
6                   & w/o decoupling  & \textbf{0.395}  & \textbf{0.122}     & 38.63          & 0.969          & 8.83           & 0.66     & 39.32 & 0.936 & 5.97  & 0.60       \\ 
\midrule
7                   & conv DE  & 0.549  & 0.154     & 39.34          & 0.974          & 7.93           & 0.56     & 40.38 & 0.940 & 5.75  & 0.58       \\ 
\midrule
                    & Ours     & 0.549  & 0.154     & \textbf{42.15} & \textbf{0.983} & \textbf{6.48}  & \textbf{0.42} & \textbf{44.62} & \textbf{0.991} & \textbf{3.76}  & \textbf{0.39} \\
\bottomrule[1.2pt]
\end{tabular}}
\caption{Ablation experiment results on CAVE and Harvard datasets. The best results are in \textbf{bold}, while the second-best results are in \underline{underlined}.}
\vspace{-0.5cm}
\label{table_ablation}
\end{table*}

\begin{figure*}[htb]
	\centering
	\includegraphics[width=0.96\linewidth]{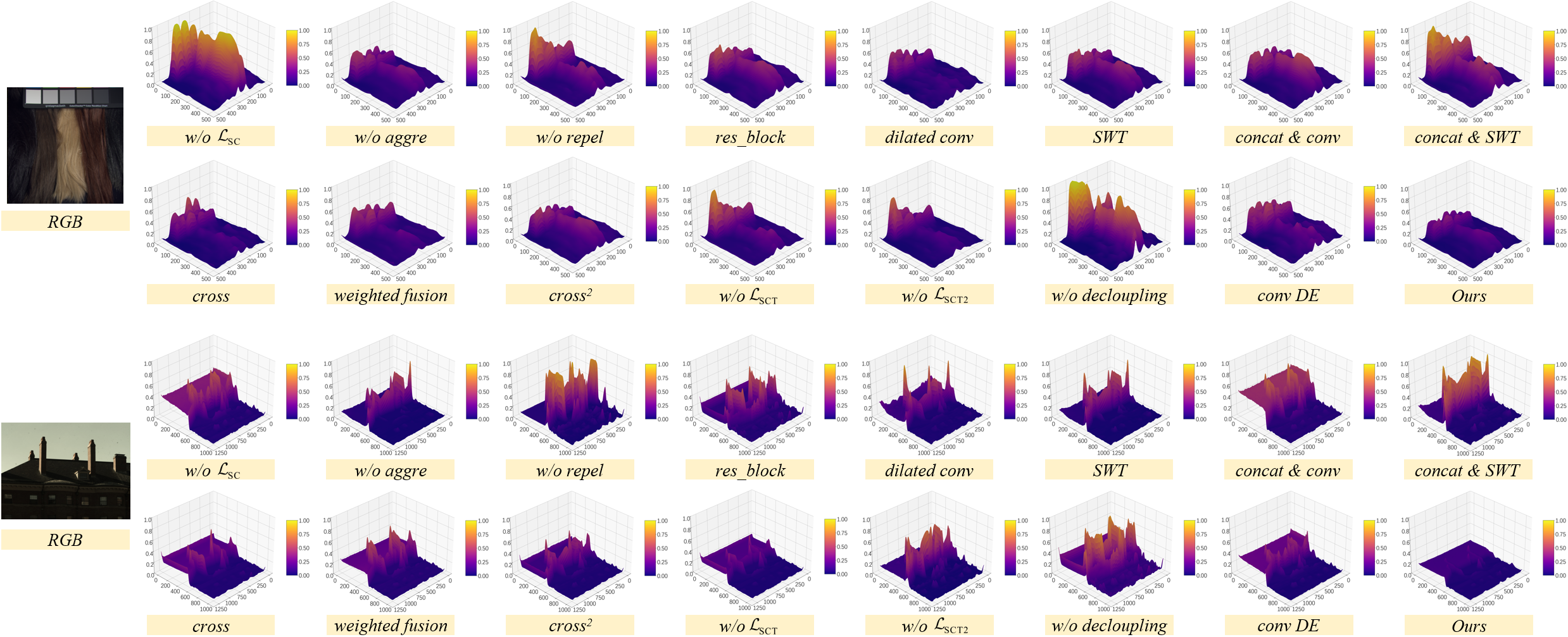}
     \caption{Visualization of cumulative reconstruction error across all spectral bands for HR-HSIs from CAVE (top) and Harvard (bottom) datasets in ablation study. Higher error values are indicated by brighter intensity, highlighting regions where fused result deviates more from ground truth.}
	\label{fig_ablation}
\end{figure*}
% \begin{figure*}
% 	\centering
% 	\includegraphics[scale=0.215]{figure/ablation_appendix.png}
% 	\caption{The reconstruction error comparison for the 11th band of the HR-HSIs from the CAVE dataset of ablation experiments.}
% 	\label{fig_ablation}
% \end{figure*}

\noindent \textbf{Ablation Study.} To rigorously validate our architectural design, we conduct comprehensive ablation studies on the CAVE and Harvard datasets, systematically evaluating four core innovation components of MossFuse: 1.Subspace Clustering Loss, 2.LK-CNN Block, 3.Spatial- and Spectral-Aware Blocks, 4.Fusion Strategy of Shared Representations, 5.Self-supervised Constraint Process, 6.Decoupling Mechanism, and 7.Degradation Estimation Process. The quantitative results, presented in Table~\ref{table_ablation}, demonstrate the contribution of each component to the overall reconstruction performance. In addition, we provide visual comparisons of the reconstruction errors in Fig.~\ref{fig_ablation}, exhibiting the cumulative error across all spectral bands. These visualizations illustrate that removing any of the core components leads to increased artifacts and degraded fidelity, particularly in texture details and spectral consistency.

\textbf{1. Subspace Clustering Loss.} To quantify the contribution of the Subspace Clustering Loss (\( \mathcal{L}_\text{SC}\)), we conduct three ablation variants: We first fully remove the \( \mathcal{L}_\text{SC}\) to assess its contribution, denoted as 'w/o \(\mathcal{L}_\text{SC}\)'; In addition, we modify the \(\mathcal{L}_\text{SC}\) by separately removing the aggregate term and the repulsion term from the loss, labeled as 'w/o aggre' and 'w/o repel,' respectively. Specifically, we reformulate \( \mathcal{L}_\text{SC}\) as the following two terms:
\begin{equation}
\mathcal{L}_\text{SC}^{\prime}=\log({f(F^{C}_{\text{Y}}, F^{C}_{x})+\sum_{\scriptscriptstyle i \in \{\text{Y}, x\}}f(F^{S}_{i}, F^{C}_{i})},
\end{equation}

\begin{equation}
\mathcal{L}_\text{SC}^{\prime\prime}=-\log(f(F^{S}_{\text{Y}}, F^{S}_{x})),
\end{equation}
thereby selectively disabling modality-shared features alignment or modality-complementary features separation. As shown in Table~\ref{table_ablation} and Fig.~\ref{fig_ablation}, all variants exhibit significant performance degradation, validating the necessity of both mechanisms. The decline in 'w/o aggre' underscores the importance of enforcing consistency between shared features, while the drop in 'w/o repel' reveals the risk of modality decoupling without explicit modality discrimination. These results confirm that \( \mathcal{L}_\text{SC}\) optimally balances 'aggregation' and 'repulsion' to enhance representation disentanglement.

\textbf{2. LK-CNN Block.} 
To address the large spatial resolution gap (up to $\times 32$) and effectively extract high-frequency spatial details, we employ the lightweight Large-Kernel CNN (LK-CNN)~\cite{zhang2024scaling}. It consists of three depthwise-separable $3\times3$ convolutions with batch normalization and ReLU activations, achieving a $7\times7$ receptive field while maintaining efficiency. To validate its effectiveness, we replaced it with three alternatives: a standard Residual Block, a stack of three dilated convolution layers ('dilated conv'), and an SWT block, denoting these variants as 'res\_block', 'dilated conv', and 'SWT', respectively. As shown in Table~\ref{table_ablation} labeled '2', all three alternatives result in noticeable performance degradation. In particular, the Residual Block is limited in both feature extraction capacity and efficiency, while dilated convolutions, although more efficient, fail to match the overall effectiveness of the LK‑CNN design. These results justify the use of LK‑CNN blocks as an effective and balanced spatial feature extractor.

\textbf{3. Spatial- and Spectral-Aware Aggregation Blocks.} To evaluate the necessity of our spatial-and spectral-aware aggregation blocks, we replace them with three baseline fusion strategies: concatenation followed by a 3\(\times\)3 convolution, concatenation followed by an SWT block and cross-attention, denoted as 'concat$\&$conv', 'concat$\&$SWT', and 'cross', respectively. As shown in Table~\ref{table_ablation}, both substitutions lead to significant performance degradation (e.g., -3.23 dB PSNR for 'concat$\&$conv', 3.13dB for 'concat$\&$SWT', and -1.86 dB for 'cross'), revealing the critical limitations of naive fusion approaches. The performance gap further demonstrates that effective multimodal fusion requires interaction from both spatial and spectral dimensions, rather than treating modalities as homogeneous inputs.

\textbf{4. Fusion Strategy of Shared Representations.} To validate the efficiency of the current fusion scheme for the two shared representations \(F^{S}_{\text{Y}}\) and \(F^{S}_{x}\), we compare with two alternative fusion strategies, including: (1) weighted fusion, which leverage a learnable scalar weight \(\omega\) to compute \(\omega \odot F^{S}_{\text{Y}} + (1 - \omega) \odot F^{S}_{x}\); (2) cross-attention-based fusion (marked as cross$^{2}$). As shown in Table~\ref{table_ablation}, both alternatives introduce additional parameters or computational overhead while yielding consistently lower fusion accuracy. In addition, the error visualization in Fig.~\ref{fig_ablation} exhibits larger reconstruction errors as compared to ours. These results demonstrate that our simple additive fusion not only reduces model complexity but also enhances performance by enabling more stable and effective aggregation of modality-shared information.

\textbf{5. Self-supervised Constraint Process.} To validate the efficacy of the self-supervised constraint mechanism, we conduct two targeted ablations: We first completely remove this process labeled as 'w/o \(\mathcal{L}_\text{MCT}\)'; In addition, we selectively remove the shared-representation supervision branch while preserving the primary constraint term of HR-MSI and LR-HSI (marked as 'w/o \(\mathcal{L}_\text{MCT2}\)'). As evidenced by the significant performance drops recorded in Table~\ref{table_ablation}, this analysis quantitatively demonstrates that the completeness and fidelity of the decoupled subspace features act as a necessary condition for model optimization. Moreover, the observed performance improvement achieved through constraining the two modality-shared features \(F^{S}_{\text{Y}}\) and \(F^{S}_{x}\) to the latent representation of LR-MSI further validates our core architectural design principle that the LR-HSI and HR-MSI can be decoupled into modality-complementary and modality-shared components that collectively characterize the latent representations of LR-MSI embedding.

\textbf{6. Decoupling Mechanism.} To comprehensively validate the effectiveness of our proposed decoupling mechanism, we carry out ablation studies beyond the investigation into the loss functions (e.g., \(\mathcal{L}_{SC}\), \(\mathcal{L}_{SCT}\)), to also consider the impact of decoupling from an architectural standpoint. The case is therefore marked as 'w/o decoupling', by removing the dedicated modality-shared encoder and modality-complementary encoders, while adopting a single encoder for each modality (namely, one for HR-MSI and one for the upsampled LR-HSI). These unified encoders directly generate modality-specific features without enforcing any explicit decomposition into shared and complementary components. These features are then concatenated and fed into a shallow \(1\times 1\) convolutional fusion block, followed by the decoder to reconstruct the final HR-HSI. This baseline closely resembles typical encoder-fusion-decoder architectures used in existing fusion methods, thus providing a fair comparison. As evidenced by the significant performance drops recorded in Table~\ref{table_ablation} and the increased reconstruction errors in Fig.~\ref{fig_ablation}, both quantitative and visual analyses confirm the effectiveness of the decoupling mechanism.

\textbf{7. Degradation Estimation Process.} The physics-driven degradation estimation process serves as a foundational component for solving the ill-posed hyperspectral inverse problem by explicitly modeling the SRF and PSF. This systematic parameter estimation enables physically meaningful constraints for both the final reconstructed hyperspectral image and the intermediate LR-MSI generation. To assess the effectiveness of our approach, we conduct ablation studies by replacing this principled estimation framework with unconstrained conventional convolutional layers (denoted as 'conv DE'). As demonstrated quantitatively in Table~\ref{table_ablation}, and qualitatively through error distribution analysis in Fig.~\ref{fig_ablation},
this substitution leads to significant performance degradation, with error heatmaps revealing amplified spectral distortions in high-frequency regions. These comparative results provide quantitative and visual confirmation of our degradation estimation module's critical role in maintaining spectral-spatial consistency throughout the reconstruction pipeline.

\section{Conclusion}
In this paper, we have presented a fundamental insight into the fusion-based hyperspectral image super-resolution task by revealing the critical importance of modality decoupling. Specifically, we have introduced an end-to-end self-supervised modality-decoupled spatial-spectral fusion (MossFuse) framework that decouples shared and complementary information across modalities and aggregates a concise representation of the LR-HSIs and HR-MSIs to reduce modality redundancy. To reinforce this process, we have developed a novel subspace clustering loss that simultaneously guides the network in aggregating shared LR-MSI components while discerning them from the modality-complementary spatial and spectral components. Through extensive experimentation, we have validated that the integration of precisely decoupled components: the modality-shared LR-MSIs and modality-complementary spatial, spectral information, substantially enhances fusion performance. This improvement is achieved without resorting to complex architectural designs or extensive parameterization. As such, this work results in superior-quality HR-HSI reconstruction with significantly reduced computational overhead in terms of both model parameters and inference time. For future work, we will explore the integration of large-scale pretraining and semantic supervision to further enhance the generalizability of our framework across diverse scenes and sensor domains. Further work will also include experimental investigations into more forms of noise distribution, which would help reveal the limitations of the present work in more effectively dealing with unseen signal degradation types.

% In this paper, we reveal that modality decoupling is essential for hyperspectral and multispectral image fusion. Specifically, we propose a simple framework that decouples the modality-shared/complementary components and aggregates the concise representation of the LR-HSI and HR-MSI to reduce the modality redundancy. Also, we proposed a subspace clustering loss to guide the network to aggregate the shared information while distinguishing it from its modality-complementary components. Extensive experiments prove that fusing the decoupled modality-shared LR-MSI and modality-complementary spatial and spectral information enhances fusion performance without the need for complex architecture or vast parameters, leading to a finer HR-HSI with fewer parameters and inference time. 

% \bmhead{Acknowledgements}
\section{Acknowledgements}
This research is supported by National Natural Science Foundation of China (62271400).

\section{Data Availability Statement}
All code and data supporting the findings and experiments of this paper are publicly available in the GitHub repository at \url{https://github.com/dusongcheng/MossFuse}.
% \section*{Declarations}

% Some journals require declarations to be submitted in a standardised format. Please check the Instructions for Authors of the journal to which you are submitting to see if you need to complete this section. If yes, your manuscript must contain the following sections under the heading `Declarations':

% \begin{itemize}
% \item Conflict of interest/Competing interests (check journal-specific guidelines for which heading to use)
% \item Ethics approval and consent to participate
% \item Consent for publication
% \item Materials availability
% \end{itemize}

% \noindent
% If any of the sections are not relevant to your manuscript, please include the heading and write `Not applicable' for that section. 

\bibliography{sn-bibliography}% common bib file

\end{document}